\documentclass[prd, notitlepage, twocolumn, superscriptaddress,nofootinbib, floatfix, longbibliography]{revtex4-2}

\pdfoutput=1

\usepackage[pdftex]{graphicx}              
\usepackage{amsmath}
\usepackage{amssymb}
\usepackage{amsfonts}

\usepackage[T1]{fontenc}
\usepackage[utf8]{inputenc}
\usepackage{microtype}

\usepackage[normalem]{ulem} %

\usepackage{anyfontsize}

\usepackage[linktocpage,breaklinks]{hyperref}
\usepackage{url}
\usepackage[all]{hypcap}
\usepackage[usenames,dvipsnames]{xcolor}

\usepackage{tensor}
\usepackage{bm}

\usepackage{graphicx}
\graphicspath{{./figures/}}

\usepackage{tikz}
\ifdefined\myext
\usetikzlibrary{external}
\fi

\usepackage{natbib}
\usepackage{hyperref}
\usepackage{orcidlink}
\usepackage{verbatim}

\hypersetup{colorlinks=true,
            citecolor=RoyalBlue,
            linkcolor=RoyalBlue,
            urlcolor=RoyalBlue}

\usepackage{xspace}
\usepackage{dsfont}
\usepackage{soul} 
\usepackage{dsfont}
\usepackage{multirow}
\usepackage{float}
\usepackage{cancel}

\usepackage[caption=false]{subfig}

\newcommand{\orcid}[1]{\href{https://orcid.org/#1}{\includegraphics[height=\fontcharht\font`\B]{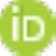}}}

\usepackage{stackengine}
\usepackage{scalerel}

\stackMath
\def\dyhat{.067ex}
\newcommand\myhat[1]{\ThisStyle{%
              \stackon[\dyhat]{\SavedStyle#1}
                      {\SavedStyle\mathbf{\hat{\phantom{#1}}}}}}

\newcommand{\beq}{\begin{equation}}
\newcommand{\eeq}{\end{equation}}

\newcommand{\nrpyell}{{\texttt{NRPyElliptic}}\xspace}
\newcommand{\nrpyellet}{{\texttt{NRPyEllipticET}}\xspace}

\newcommand{\SinhSymTP}{{\texttt{SinhSymTP}}\xspace}
\newcommand{\nrpy}{{\texttt{NRPy+}}\xspace}
\newcommand{\TwoPunctures}{{\texttt{TwoPunctures}}\xspace}
\newcommand{\etk}{\texttt{Einstein Toolkit}\xspace}
\newcommand{\newrad}{\texttt{NewRad}\xspace}
\newcommand{\SymPy}{\texttt{SymPy}\xspace}
\newcommand{\Jupyter}{\texttt{Jupyter}\xspace}
\newcommand{\bhah}{{\texttt{BlackHoles@Home}}\xspace}

\newcommand{\that}[1]{\mathbf{\myhat{\text{$#1$}}}}

\newcommand{\ttilde}[1]{\mathbf{\tilde{\text{$#1$}}}}

\newcommand{\xxzero}{\mathtt{x_{1}}}   
\newcommand{\xxone}{\mathtt{x_{2}}}    
\newcommand{\xxtwo}{\mathtt{x_{3}}}    
\newcommand{\dxxzero}{\mathtt{\Delta x_{1}}} 
\newcommand{\dxxone}{\mathtt{\Delta x_{2}}}  
\newcommand{\dxxtwo}{\mathtt{\Delta x_{3}}}  
\newcommand{\dxxi}{\mathtt{\Delta x_{i}}}  
\newcommand{\Nzero}{\mathtt{N_{1}}}    
\newcommand{\None}{\mathtt{N_{2}}}     
\newcommand{\Ntwo}{\mathtt{N_{3}}}     
\newcommand{\ione}{\mathtt{i_{2}}}     
\newcommand{\hone}{h^{(2)}}          
\newcommand{\htwo}{h^{(3)}}          
\newcommand{\sinhA}{\mathcal{A}}    
\newcommand{\sinhW}{w}              
\newcommand{\bScale}{b}             
\newcommand{\dt}{\Delta t}          
\newcommand{\ds}{\Delta s}          
\newcommand{\dS}{\Delta S}          
\newcommand{\dSone}{\dS^{(2)}}       
\newcommand{\dStwo}{\dS^{(3)}}       
\renewcommand{\eqref}[1]{Eq.\,(\ref{#1})}
\newcommand{\eqsref}[1]{Eqs.\,(\ref{#1})}
\newcommand{\eqrefalt}[1]{Eq.\,\ref{#1}}
\newcommand{\eqsrefalt}[1]{Eqs.\,\ref{#1}}

\DeclareMathOperator\arctanh{arctanh}

\newcommand{\wavespeed}{c}
\newcommand{\ellop}{\mathcal{L}_{E}}

\definecolor{amethyst}{rgb}{0.6, 0.4, 0.8}

\begin{document}
\date{\today}
\title{\nrpyell: A Fast Hyperbolic Relaxation Elliptic
  Solver for Numerical Relativity, I: Conformally Flat, Binary Puncture
  Initial Data}
\author{Thiago Assump\c c\~ao~\orcid{0000-0002-3419-892X}}
\email{ta0082@mix.wvu.edu}
\affiliation{Department of Physics and Astronomy, West Virginia University, Morgantown, WV 26506, USA}
\affiliation{Center for Gravitational Waves and Cosmology, West Virginia University, Chestnut Ridge Research Building, Morgantown, WV 26505, USA}

\author{Leonardo R. Werneck~\orcid{0000-0002-4541-8553}}
\email{leonardo@uidaho.edu}
\affiliation{Department of Physics, University of Idaho, Moscow, ID 83843, USA}
\affiliation{Department of Physics and Astronomy, West Virginia University, Morgantown, WV 26506, USA}
\affiliation{Center for Gravitational Waves and Cosmology, West Virginia University, Chestnut Ridge Research Building, Morgantown, WV 26505, USA}

\author{Terrence Pierre Jacques~\orcid{0000-0002-8993-0567}}
\email{tp0052@mix.wvu.edu}
\affiliation{Department of Physics and Astronomy, West Virginia University, Morgantown, WV 26506, USA}
\affiliation{Center for Gravitational Waves and Cosmology, West Virginia University, Chestnut Ridge Research Building, Morgantown, WV 26505, USA}

\author{Zachariah B.~Etienne~\orcid{0000-0002-6838-9185}}
\email{zetienne@uidaho.edu}
\affiliation{Department of Physics, University of Idaho, Moscow, ID 83843, USA}
\affiliation{Department of Physics and Astronomy, West Virginia University, Morgantown, WV 26506, USA}
\affiliation{Center for Gravitational Waves and Cosmology, West Virginia University, Chestnut Ridge Research Building, Morgantown, WV 26505, USA}

\begin{abstract}
We introduce \nrpyell, an elliptic solver for numerical relativity (NR)
built within the \nrpy framework. As its first application, \nrpyell
sets up conformally flat, binary black hole (BBH) puncture initial
data (ID) on a single numerical domain, similar to the widely used
\TwoPunctures code. Unlike \TwoPunctures,
\nrpyell employs a hyperbolic relaxation scheme, whereby arbitrary
elliptic PDEs are trivially transformed into a hyperbolic system of
PDEs. As consumers of NR ID generally already possess expertise in
solving hyperbolic PDEs, they will generally find \nrpyell easier
to tweak and extend than other NR elliptic solvers. When evolved
forward in (pseudo)time, the hyperbolic system exponentially reaches a
steady state that solves the elliptic PDEs. Notably \nrpyell
accelerates the relaxation waves, which makes it many orders of
magnitude faster than the usual constant-wavespeed approach. While it
is still ${\sim}12$x slower than \TwoPunctures at setting up full-3D
BBH ID, \nrpyell requires only ${\approx}0.3\%$
of the runtime for a full BBH simulation in the \etk. Future work will
focus on improving performance and generating other types of ID, such
as binary neutron star.
\end{abstract}

\maketitle

\section{Introduction}
\label{sec:intro}

To date the LIGO/Virgo gravitational wave (GW) observatories have
detected dozens of binary black hole (BBH)
mergers~\cite{LIGOScientific:2021usb}, and numerical relativity (NR)
BBH simulations form a cornerstone of the ensuing data
analyses. Such simulations build from
formulations
of the general relativistic (GR) field equations~\cite{Nakamura:1987zz,Shibata:1995we,Baumgarte:1998te,Pretorius:2004jg,Bona:2003fj,Bernuzzi:2009ex,Alic:2011gg,Alic:2013xsa,Sanchis-Gual:2014nha} that decompose GR
into an initial value problem in 3+1 dimensions.
These formulations generally rewrite the GR field equations as a set
of time evolution and constraint equations, similar to
Maxwell's equations in differential form~\cite{Knapp:2002fm}. Thus, so
long as one is provided initial data (ID) that satisfy the Einstein
constraints (elliptic PDEs), the evolution equations (hyperbolic PDEs) can propagate them forward in
time to construct the spacetime.


Construction of NR ID for realistic
astrophysical scenarios is typically a complex and highly
specialized task (see~\cite{Cook:2000vr,Pfeiffer:2005jhde,Tichy_2016,Gourgoulhon:2007tn} for excellent reviews). A
wide range of approaches have been employed by different groups to set
up ID for realistic astrophysical scenarios,
including the use of finite
difference~\cite{Tsokaros:2007,Uryu:2009,Tsokaros:2015fea,Uryu:2012,East:2012,Moldenhauer:2014},
spectral~\cite{Grandclement:2006,Gourgoulhon:2000nn,Pfeiffer:2002wt,Ossokine:2015,Foucart:2008,Tacik:2015,Tacik:2016,Taniguchi:2001qv,Taniguchi:2002ns,Ansorg:2004ds,Dietrich:2015mjb,Wolfgang:2019,Grandclement:2010p,Papenfort:2021,Rashti:2021},
and Galerkin~\cite{Barreto:2018,Vu:2021coj} methods, which are then combined with
special numerical techniques to solve the associated elliptic PDEs. Notably, this generally requires a different skill set than those
associated with solving the (hyperbolic) evolution equations,
so experts in setting up ID for NR are rarely
experts in solving the evolution equations, and {\it vice versa}.

\nrpyell is a new, extensible elliptic solver that sets up initial data for numerical relativity using the same numerical methods employed for
solving hyperbolic evolution equations. Specifically, \nrpyell
implements the hyperbolic relaxation method of~\cite{Ruter:2017iph}
to solve complex nonlinear elliptic PDEs for NR ID. The hyperbolic PDEs are evolved forward in
(pseudo)time, resulting in an exponential relaxation of the arbitrary initial
guess to a steady state that coincides with the solution of
the elliptic system. \nrpyell solves these
equations on highly
efficient numerical grids exploiting underlying symmetries in the
physical scenario. To this end, \nrpyell is built within the
\SymPy~\cite{meurer2017sympy}-based \nrpy code-generation
framework~\cite{Ruchlin:2017com,nrpy_web}, which facilitates the
solution of hyperbolic PDEs on Cartesian-like, spherical-like,
cylindrical-like, or bispherical-like numerical grids. For the
purposes of setting up BBH puncture ID, \nrpyell makes use of the
latter.

Choice of appropriate numerical grids is critically important, as
setting up binary compact object ID requires numerically resolving many orders of magnitude in length scale: from the sharp
gravitational fields near each compact object, to the nearly flat
fields far away. If a constant wavespeed is chosen in
the hyperbolic relaxation method (as in~\cite{Ruter:2017iph}), the
Courant-Friedrichs-Lewy (CFL)-constrained global timestep for the relaxation will be many
orders of magnitude smaller than the time required for a relaxation
wave to cross the numerical domain. This poses a significant
problem as hyperbolic relaxation methods must propagate the relaxation
waves across the {\it entire} numerical domain {\it several times} to reach
convergence. Thus hyperbolic relaxation solvers are generally
far slower than elliptic solvers based on specialized numerical
methods.

\nrpyell solves the hyperbolic relaxation PDEs on a single
bispherical-like domain, enabling us to increase the local relaxation
wavespeed in proportion to the local grid spacing without violating
the CFL condition. As grid spacing in our coordinate system grows
{\it exponentially} away from the two coordinate foci and toward the outer boundary,
the relaxation waves accelerate exponentially toward the outer
boundary, increasing the solver's overall speed-up over a
constant-wavespeed implementation by many orders of magnitude.
In fact the resulting performance boost enables \nrpyell to be useful
for setting up high-quality, full-3D BBH puncture ID, though it is
still 12x slower than the widely used pseudospectral
\TwoPunctures~\cite{Ansorg:2004ds} BBH puncture ID solver.\footnote{Benchmark tests performed using an Intel Xeon Gold 6230 20-Core CPU for initial data of comparable quality.}

Like \TwoPunctures, \nrpyell adopts the conformal transverse-traceless
decomposition~\cite{York:1998hy,Cook:2000vr,Lovelace:2011nu,Alcubierre08}
to construct puncture ID for two BHs. In this paper we present both 2D and full 3D
validation tests, which demonstrate that \nrpyell yields identical
results to \TwoPunctures as numerical resolution is increased in
both codes.

We also embed \nrpyell into an
\etk~\cite{Loffler:2011ay,etk_web,Cactusweb} module (``thorn''), called \nrpyellet,
which enables the generated ID to be interpolated onto Cartesian AMR (adaptive mesh refinement)
grids within the \etk. To demonstrate that \nrpyell ID are of high
fidelity, we first generate 3D BBH puncture ID with both \nrpyellet and the
\TwoPunctures \etk thorns at comparable-accuracy; then evolve the ID forward in time through
inspiral, merger, and ringdown using the \etk infrastructure; and finally show that the results of
these simulations are virtually indistinguishable.

While existing algorithms for elliptic equations can be used to solve a general class of problems, their numerical implementations can be rather problem-specific (see, e.g., Ref.~\cite{Press92} and references therein). For instance, the \TwoPunctures code employs the Biconjugate Gradient Stabilized (BiCSTAB) method, which requires a specific preconditioner for numerical efficiency~\cite{Ansorg:2004ds}. In contrast, hyperbolic relaxation solvers can be trivially used for other elliptic problems by simply replacing the right-hand sides of the resulting hyperbolic PDEs while preserving other aspects of the numerical implementation. In fact the generality of the hyperbolic relaxation method has already been
demonstrated in~\cite{Ruter:2017iph}, where it was used to
produce ID for many different scenarios of interest, such as
scalar fields, Tolman-Oppenheimer-Volkoff (TOV) stars, and binary
neutron stars (BNSs). In this work we will focus our discussion on BBH
puncture ID, and further evidence of the extensibility of \nrpyell
will be presented in forthcoming papers to generate e.g., BNS ID.

The remainder of this paper is organized as follows. Sec.~\ref{sec:basic_equations}
introduces the puncture ID formalism, the hyperbolic relaxation method,
and our implementation of Sommerfeld (radiation) boundary conditions. In
Sec.~\ref{sec:numerical_implementation} we discuss the details of our
numerical implementation, including choice of coordinate system and
implementation of a grid spacing-dependent wavespeed. We present
2D (axisymmetric) and full 3D validation tests, as well as results
from a BBH evolution of our full 3D ID in
Sec.~\ref{sec:results}. We conclude in Sec.~\ref{sec:conclusions} and
discuss future work.

\section{Basic equations}
\label{sec:basic_equations}
Throughout this paper we adopt geometrized units, in which $G=c=1$, and
Einstein summation convention such that repeated Latin (Greek) indices
imply a sum over all 3 spatial (all 4 spacetime) components.

Consider the $3+1$ decomposition of the spacetime metric, with line element
\begin{equation}
ds^2 = -\alpha^2 dt^2 + \gamma_{ij}(dx^i + \beta^i dt) (dx^j + \beta^j dt)\,.
\end{equation}
Here, $\alpha$ is the lapse function, $\beta^{i}$ is the shift vector,
and $\gamma_{ij}$ is the $3$-metric.

It is useful to define a
conformally related $3$-metric $\ttilde{\gamma}_{ij}$ via
\begin{equation}
\gamma_{ij} = \psi^4 \ttilde{\gamma}_{ij} \,,
\end{equation}
where the scalar function $\psi$ is known as the conformal
factor. We adorn geometric quantities associated with
$\ttilde{\gamma}_{ij}$ with a tilde diacritic. For instance, the Christoffel
symbols associated with $\ttilde{\gamma}_{ij}$ are computed using
\begin{equation}
  \ttilde\Gamma^{k}_{\ i j} = \frac{1}{2} \ttilde\gamma^{l k} (\ttilde\gamma_{l i, j} +
                                                           \ttilde\gamma_{l j, i} -
                                                           \ttilde\gamma_{i j, l} ) \,.
\end{equation}
Likewise, $\ttilde{\nabla}_{i}$ is the associated conformal covariant
derivative and $\ttilde{R}_{ij}$ the Ricci tensor. All geometric
quantities compatible with the physical $3$-metric $\gamma_{ij}$ are
written without tildes.

In the limit of vacuum (e.g., BBH) spacetimes, the Hamiltonian
and momentum constraint equations can be written as~\cite{Cook:2000vr}
\begin{align}
\label{eqn:hamiltonian_constraint}
\mathcal{H} &\equiv R + K^2 - K_{ij} K^{ij} = 0 \,, \\
\label{eqn:momentum_constraint}
\mathcal{M}^{i} &\equiv \nabla_{j}(K^{ij} - \gamma^{ij}K) = 0 \,,
\end{align}
where $K_{ij}$ is the extrinsic curvature and $K\equiv\gamma^{ij}K_{ij}$
is the mean curvature. Setting up ID for vacuum spacetimes in numerical
relativity generally involves solving these constraints, which exist as second-order nonlinear
elliptic PDEs.

For the purposes of this paper, we will focus on the puncture ID
formalism, in which a set of simplifying assumptions is applied to these
constraints, known as the conformal transverse-traceless (CTT) decomposition
(see e.g.,~\cite{Cook:2000vr}). For completeness we next apply the CTT
approach to
Eqs.~(\ref{eqn:hamiltonian_constraint},~\ref{eqn:momentum_constraint})
to derive the constraint equations solved in this paper by \nrpyell.

\subsection{Puncture Initial Data Formalism}
\label{sec:initial_data}

To arrive at the CTT decomposition, we first rewrite the extrinsic
curvature as
\begin{equation}
\label{eqn:tracefree_K}
K_{ij} = A_{ij} + \frac{1}{3} \gamma_{ij} K \,,
\end{equation}
where $A_{ij}$ is the trace-free part of $K_{ij}$. The conformal
counterpart of $A_{ij}$ is defined through the relation
\begin{equation}
A_{ij} \equiv \psi^{-2} \ttilde{A}_{ij} \,.
\end{equation}

The CTT decomposition splits
$\ttilde{A}_{ij}$ into a symmetric trace-free part $\ttilde{M}^{ij}$
and a longitudinal part $(\ttilde{\mathds{L}} V)^{ij}$,
\begin{equation}
\ttilde{A}^{ij} = (\ttilde{\mathds{L}} V)^{ij} + \ttilde{M}^{ij} \,,
\end{equation}
where the longitudinal operator $\ttilde{\mathds{L}}$ is defined via
\begin{equation}
(\ttilde{\mathds{L}} V)^{ij} \equiv \ttilde{\nabla}^{i} V^{j} + \ttilde{\nabla}^{j} V^{i} \
                                 - \frac{2}{3} \ttilde{\gamma}^{ij} \ttilde{\nabla}_{l} V^{l} \,.
\end{equation}
Inserting these CTT quantities into the constraint equations
(Eqs.~\ref{eqn:hamiltonian_constraint},~\ref{eqn:momentum_constraint}) yields
the generic CTT Hamiltonian and momentum constraint equations (see, e.g.,~\cite{BaumgarteShapiro_2010,Alcubierre08} and references therein)
\begin{align}
\ttilde{\nabla}^2 \psi - \frac{1}{8}\psi \ttilde{R} - \frac{1}{12}\psi^5 K^2 + \frac{1}{8}\psi^{-7} \ttilde{A}_{ij} \ttilde{A}^{ij} &= 0 \,, \\
\ttilde{\Delta}_{\mathds{L}} V^i - \frac{2}{3} \psi^6 \ttilde{\nabla}^i K + \ttilde{\nabla}_j \ttilde{M}^{ij} &= 0 \,,\label{eq:MU_cttd}
\end{align}
where $\ttilde{R}$ is the conformal Ricci scalar and the operator $\ttilde{\Delta}_{\mathds{L}}$ is defined as
\begin{equation}
\ttilde{\Delta}_{\mathds{L}} V^i \equiv \ttilde{\nabla}_j (\ttilde{\mathds{L}} V)^ {ij} =  \ttilde{\nabla}^2 V^i + \frac{1}{3} \ttilde{\nabla}^i(\ttilde{\nabla}_j V^j) + \ttilde{R}^{i}_{\ j} V^j \,.
\end{equation}

The degrees of freedom in this formulation include choice of $\ttilde{M}^{ij}$, $K$, and $\ttilde\gamma_{ij}$. Here we consider puncture ID, which assume maximal slicing ($K = 0$),
asymptotic flatness ($\psi|_{r\to\infty}=1$), and conformal flatness
\begin{equation}
\ttilde\gamma_{ij} = \that\gamma_{ij}\,,
  \label{eq:BY_confflatness}
\end{equation}
where $\that\gamma_{ij}$ is the flat spatial metric. In addition the assumption
$\ttilde{M}^{ij} = 0$ is made, yielding Hamiltonian and momentum constraint equations of the form~\cite{Ansorg:2004ds}
\begin{align}
\label{eqn:hamiltonian_constraint_v2}
\that{\nabla}^2 \psi + \frac{1}{8}\psi^{-7} \ttilde{A}_{ij} \ttilde{A}^{ij} &= 0 \,, \\
\label{eqn:momentum_constraint_v2}
\that{\nabla}^2 V^i + \frac{1}{3} \that{\nabla}^i (\that{\nabla}_j V^j) + \that{R}^{i}_{\ j} V^j &= 0\,,
\end{align}
where $\that\nabla_{i}$ is the covariant derivative compatible with
$\that\gamma_{ij}$. Bowen and York~\cite{Bowen:1980yu} showed that the
momentum constraint is solved
for a set of $N_p$ punctures with a
closed-form expression for the extrinsic curvature. This expression
can be written in terms of $\vec{V}$ as follows
\begin{equation}
    \vec{V} = \sum_{n=1}^{N_p}
\left(
- \frac{7}{4 |\vec{x}_n|} \vec{P}_n
- \frac{\vec{x}_n {\cdot} \vec{P}_n}{4 |\vec{x}_n|^3} \vec{x}_n
+ \frac{1}{|\vec{x}_n|^3} \vec{x}_n {\times} \vec{S}_n
\right),
\end{equation}
where $\vec{x}_n = (x_n - x, y_n - y, z_n - z)$, $\vec{P}_n$, and
$\vec{S}_n$ are the displacement relative to the origin (i.e., $(x,y,z)=(0,0,0)$), linear momentum, and
spin angular momentum of puncture $n$, respectively.

The Hamiltonian constraint equation
(Eq.~\ref{eqn:hamiltonian_constraint_v2}) must be solved numerically, but $\psi$ becoming singular at
the location of each puncture could spoil the numerical solution. Early
attempts excised the singular
terms from the computational domain (see, e.g.,~\cite{Cook:2000vr}), but modern
approaches generally follow~\cite{Brandt:1997tf} in splitting the
conformal factor into a singular and a non-singular piece,
\begin{equation}
\psi = \psi_{\rm singular} + u \equiv 1 + \sum_{n=1}^{N_p} \frac{m_{n}}{2|\vec{x}_{n} |} + u \,,
\end{equation}
where $m_n$ is the bare mass of the $n^{\rm{th}}$ puncture. The
Hamiltonian constraint equation, which can then be solved for the
non-singular part $u$, reads
\begin{equation}
\label{eqn:hamiltonian_constraint_v3}
\that{\nabla}^2 u + \frac{1}{8} \ttilde{A}_{ij} \ttilde{A}^{ij} (\psi_{\rm singular} +  u)^{-7} = 0 \,,
\end{equation}
since the Laplacian of the singular piece vanishes.

\subsection{Hyperbolic relaxation method}
\label{sec:hyperbolic_relaxation_method}

We now describe the basic hyperbolic relaxation method
of~\cite{Ruter:2017iph}. Consider the system of elliptic
equations
\begin{equation}
  \ellop \vec{u} - \vec{\rho} = 0\,,\label{eq:elliptic_equation}
\end{equation}
where $\ellop$ is an elliptic operator, $\vec{u}$ is the vector of
unknowns, and $\vec{\rho}$ is
the vector of source terms. The hyperbolic relaxation
method replaces~\eqref{eq:elliptic_equation} with the
hyperbolic system of equations
\begin{equation}
  \partial_{t}^{2}\vec{u} + \eta\partial_{t}\vec{u} = \wavespeed^{2}\bigl(\ellop \vec{u} - \vec{\rho}\bigr)\,,\label{eq:hyperbolic_relaxation_prototype}
\end{equation}
where $\eta$ is an exponential damping parameter (with units of
$1/t$~\cite{Schnetter:2010cz}) and $\wavespeed$ is the wavespeed. The
variable $t$ behaves as a time variable in this hyperbolic system of
equations and is referred to as a {\it relaxation} (as opposed to
physical) time. As noted in~\cite{Ruter:2017iph}, the damping
parameter $\eta$ that maximizes dissipation is dictated by the length
scale of the grid domain when the wavespeed
is constant. On the other hand, a spatially varying wavespeed, as introduced in
Sec.~\ref{sec:wavespeed}, gives rise to
a new scale to the problem: the relaxation-wave-crossing time, $T_{\rm{RC}}$ (refer to
Appendix~\ref{sec:appendix_wavespeed} for details). Through numerical experimentation, we
found that the choice $\eta$ that minimizes the required relaxation time follows a power law given by~\eqref{eqn:eta_power_law}.

If appropriate boundary conditions are chosen, when
\eqref{eq:hyperbolic_relaxation_prototype} is evolved forward in
(pseudo)time, the damping ensures that a steady state is eventually
reached exponentially fast such that $\partial_{t}u\to 0$ and
$\partial_{t}^{2}u \to 0$. Thus $u$ relaxes to a solution to the
original elliptic problem.
To this end, we adopt Sommerfeld (outgoing radiation) boundary conditions (BCs) for spatial boundaries, as
described in Sec.~\ref{sec:boundary_conditions}; what remains is a
choice of initial conditions. As this is a relaxation method, any
smooth choice should suffice. For simplicity, in this work we set
trivial initial conditions $\vec{u}=\partial_t\vec{u}=\vec{0}$.

To complete our expression of these equations in preparation for a
full numerical implementation, we rewrite
\eqref{eq:hyperbolic_relaxation_prototype} as a set of two
first-order (in time) PDEs
\begin{equation}
  \label{eq:generic_hyprelaxation}
  \begin{split}
    \partial_{t}\vec{u} &= \vec{v} - \eta \vec{u}\,,\\
    \partial_{t}\vec{v} &= \wavespeed^{2}\bigl(\ellop \vec{u} - \vec{\rho}\bigr)\,,
  \end{split}
\end{equation}
so that the method of lines (Sec.~\ref{sec:numerical_implementation})
can be immediately used to propagate the solution forward in
(pseudo)time until a convergence criterion has been triggered
(indicating numerical errors associated with the solution to the
elliptic equation are satisfactorily small).

As a simple example, consider Poisson's equation, for which $\ellop
\vec{u}=\ellop u = \nabla^{2} u = u^{,i}_{\ ,i}$. This PDE
can be easily made covariant (``comma goes to semicolon rule''):
\begin{equation}
  \that\nabla^{2}u = u^{;i}_{\ ;i} = \rho\,,\label{eq:poisson_rfm}
\end{equation}
where $\that\nabla_{i}$ is the covariant derivative compatible with
$\that\gamma_{ij}$. In this way, the Laplace operator is expanded as
\begin{equation}
  \that\nabla^{i}\that\nabla_{i}u = \that\gamma^{ij}\that\nabla_{i}\that\nabla_{j}u = \that\gamma^{ij}\bigl(\partial_{i}\partial_{j}u - \that\Gamma^{k}_{\ ij}\partial_{k}u\bigr)\,,
  \label{eq:laplace_operator}
\end{equation}
with $\that\Gamma^{k}_{\ ij}$ the Christoffel symbols associated with
$\that\gamma_{ij}$. Poisson's equation is then written as the system
\begin{equation}
  \boxed{
    \begin{aligned}
      \partial_{t}u &= v - \eta u\\
      \partial_{t}v &= \wavespeed^{2}\bigl(\that\nabla^{2}u - \rho\bigr)
    \end{aligned}
  }\ .\label{eq:poisson_rfm_system}
\end{equation}

Writing the PDEs covariantly enables the hyperbolic relaxation
method to be applied in coordinate systems that
properly exploit near-symmetries. For this purpose we adopt a reference metric
$\that\gamma_{ij}$, which is chosen to be the flat spatial metric in
the given coordinate system we are using. In this way, single compact object
ID can be solved in spherical or cylindrical coordinates
(using spherical or cylindrical reference metrics respectively), and
binary ID can be solved in bispherical-like coordinates. 
Once the appropriate coordinate system is chosen, all covariant derivatives are expanded
in terms of partial derivatives and the Christoffel symbols as prescribed
in~\eqref{eq:laplace_operator}.

Truly the power of the hyperbolic relaxation method is its easy and
immediate extension to complex, nonlinear elliptic PDEs by simply modifying the right-hand sides of \eqsref{eq:poisson_rfm_system}. Case in
point: ID for two punctures are constructed by solving ~\eqref{eqn:hamiltonian_constraint_v3}
for $u$. This elliptic PDE is nonlinear, but is trivially embedded within the hyperbolic relaxation prescription via\footnote{Recall in the previous section we adopted the
tilde for the conformal metric and the hat diacritic to denote the flat metric, consistent
with the general convention in the literature. Due to the choice of conformal flatness, both can be used interchangeably here, i.e., $\ttilde\gamma_{ij} = \that\gamma_{ij}$.}
\begin{equation}
  \boxed{
    \begin{aligned}
      \partial_{t}u &= v - \eta u\\
      \partial_{t}v &= c^{2}\biggl[\ttilde\nabla^{2}u + \frac{1}{8}\ttilde{A}_{ij}\ttilde{A}^{ij}\bigl(\psi_{\rm singular}+u\bigr)^{-7}\biggr]
    \end{aligned}
  }\ .\label{eq:two_punctures_rfm_system}
\end{equation}
Note that just like in the case of Poisson's equation,
$\ttilde\nabla^{2}u$ is expanded as in~\eqref{eq:laplace_operator}.

\subsection{Boundary conditions}
\label{sec:boundary_conditions}

Similar to both the hyperbolic relaxation method implemented
in~\cite{Ruter:2017iph} and the \etk BC driver
\newrad~\cite{Alcubierre:2002kk,Loffler:2011ay,etk_web}, spatial BCs
are applied to the time derivatives of the evolved fields instead of
the fields directly. Consequently the desired BC is only
satisfied by the steady state solution.

For example, assume that at
$\partial\Omega$, the boundary of our numerical domain, we wish to
impose Dirichlet BCs of the form
\begin{equation}
  \begin{split}
    \bigl.\vec{u}\bigr|_{\partial\Omega}&=\vec{a}\,,\\
    \bigl.\vec{v}\bigr|_{\partial\Omega}&=\vec{b}\,,
  \end{split}
\end{equation}
for some constant vectors $\vec{a}$ and $\vec{b}$. In our implementation, these would be imposed as
\begin{equation}
  \begin{split}
    \bigl.\partial_{t}\vec{u}\bigr|_{\partial\Omega} &= \vec{u} - \vec{a}\,,\\
    \bigl.\partial_{t}\vec{v}\bigr|_{\partial\Omega} &= \vec{v} - \vec{b}\,.
  \end{split}
\end{equation}
Upon reaching the steady state,
$\bigl.\partial_{t}u\bigr|_{\partial\Omega}=0=\bigl.\partial_{t}v\bigr|_{\partial\Omega}$,
and we recover the desired BCs.

When applying the hyperbolic relaxation method to solve the Einstein
constraint equations, outgoing radiation BCs are most appropriate,
as they allow the outgoing relaxation wave fronts to pass through the
boundaries of the numerical domain with minimal reflection.

Radiation (Sommerfeld) BCs generally assume that near the boundary each field $f$
behaves as an outgoing spherical wave, and our implementation follows
the implementation within \newrad, building upon the {\it ansatz}:
\begin{equation}
  f = f_0 + \frac{w(r-ct)}{r} + \frac{C}{r^2}\,,
\end{equation}
where $f_{0} = \lim_{r\to\infty} f$, $w(r-ct)/r$ satisfies the spherical
wave equation for an outgoing spherical wave, and $C/r^2$ models
higher-order radial corrections.

Just as in the case of Dirichlet BCs, we apply Sommerfeld BCs to the
time derivative of the fields. Appendix~\ref{sec:appendix_sommerfeld}
walks through the full derivation for applying Sommerfeld BCs to any
field $\partial_t f$, as well as its numerical implementation. Based
on \eqref{eq:sommerfeld_final}, Sommerfeld BCs for a generic
hyperbolic relaxation of solution vector $\vec{u}$ takes the form
\begin{equation}
  \begin{alignedat}{3}
    \bigl.\partial_{t}\vec{u}\bigr|_{\partial\Omega} &= -\frac{\wavespeed}{r}\biggl[r\partial_r\vec{u} &+& (\vec{u} - \vec{u}_{0})&\biggr]& + \frac{\vec{k}_u}{r^{3}}\,,\\
    \bigl.\partial_{t}\vec{v}\bigr|_{\partial\Omega} &= -\frac{\wavespeed}{r}\biggl[r\partial_r\vec{v} &+& (\vec{v} - \vec{v}_{0})&\biggr]& + \frac{\vec{k}_v}{r^{3}}\,,
  \end{alignedat}
  \label{eq:hyprelax_sommerfeld}
\end{equation}
where $\vec{k}_u$ and $\vec{k}_v$ are constant vectors computed at
each boundary point for each field within the $\vec{u}$ and $\vec{v}$
vectors using \eqref{eq:sommerfeld_expression_for_k}.


\section{Numerical implementation}
\label{sec:numerical_implementation}

\nrpyell exists as both a standalone code and an \etk module (``thorn''),
\texttt{NRPyEllipticET}. \texttt{NRPyEllipticET}
incorporates the standalone code into the \etk, solving the
elliptic PDE entirely within \nrpyell's \nrpy-based
infrastructure. Once the solution has been found,
\texttt{NRPyEllipticET} uses the \etk's built-in (3\textsuperscript{rd}-order Hermite) interpolation
infrastructure to
interpolate the solution from
its native, bispherical-like grids to the Cartesian AMR grids
used by the \etk. From there, the data can be evolved forward in time
using any of the various BSSN or CCZ4 \etk thorns. Both standalone and \etk thorn versions of the \nrpyell code are
fully documented in pedagogical \Jupyter notebooks. Henceforth we will
describe our implementation of the standalone version.

Our implementation of
\eqsref{eq:two_punctures_rfm_system} within \nrpyell leverages the \nrpy
framework~\cite{Ruchlin:2017com,nrpy_web} to convert these expressions,
written symbolically using \nrpy's Einstein-like notation, into highly
optimized C-code kernels (\SymPy~\cite{meurer2017sympy} serves as
\nrpy's computer algebra system backend). Notably \nrpy
supports the generation of such kernels with single instruction,
multiple data (SIMD) intrinsics and common sub-expression elimination
(CSE). Like \TwoPunctures, \nrpyell currently supports \texttt{OpenMP} parallelization~\cite{openmp_web} and both codes run on single computational nodes. 
Further \nrpy supports arbitrary-order
finite-difference kernel generation, and we use
10\textsuperscript{th}-order to approximate all spatial derivatives in
this work. The time evolution is performed
using the method of lines (MoL) infrastructure within \nrpy, choosing
its fourth-order (explicit) Runge-Kutta implementation (RK4).

\nrpy supports a plethora of different
reference metrics, enabling us to solve our covariant hyperbolic PDEs
(\eqsrefalt{eq:two_punctures_rfm_system}) in a large variety of
Cartesian-like, spherical-like, cylindrical-like, or bispherical-like
coordinate systems. This in turn enables the user to fully
take advantage of symmetries or near-symmetries of any given problem.
For example, for problems involving near-spherical symmetry we have
used spherical-like coordinates (e.g., log-radial
spherical coordinates). In this work, we make use of the
prolate spheroidal-like (i.e., ``bispherical-like'')
coordinate system in \nrpy called \SinhSymTP, described in
detail in Sec.~\ref{sec:coordinate_system}. This allows us to solve
the elliptic problem for two puncture black hole initial data within a single domain, similar to the \TwoPunctures code.

Note that the wavespeed $c$ appearing e.g., in
\eqsref{eq:two_punctures_rfm_system}, need not be constant. In
curvilinear coordinates where the grid spacing is not constant, the
CFL stability criterion remains satisfied if the wavespeed is
adjusted in proportion to the local grid spacing. As the grid spacing
in the \SinhSymTP coordinates adopted here grows {\it exponentially}
with distance from the strong-field region, the wavespeed grows
exponentially as well. As a result, relaxation waves accelerate
exponentially to the outer boundary, significantly speeding up the
convergence to the solution of the elliptic PDE. Our implementation of
this technique is detailed in Sec.~\ref{sec:wavespeed}.

\subsection{Coordinate system}
\label{sec:coordinate_system}

Like \TwoPunctures, \nrpyell adopts a modified
version of prolate spheroidal (PS) coordinates when setting up
two-puncture ID. However, these coordinate systems are distinct both
from each other and from PS coordinates. Here we elucidate the differences and
similarities.

Consider first PS coordinates $(\mu,\nu,\varphi)$, which are related
to Cartesian coordinates $(x,y,z)$ via~\cite{moon2012field}
\begin{align}
  x &= a\sinh\mu \sin\nu \cos\varphi\,,\nonumber\\
  y &= a\sinh\mu \sin\nu \sin\varphi\,,\label{eq:PSC_xyz}\\
  z &= \left(a^{2}\sinh^{2}\mu + a^{2}\right)^{1/2}\cos\nu\,.\nonumber
\end{align}
Here, $\mu\in[0,\infty)$, $\nu\in[0,\pi)$, $\varphi\in[0,2\pi)$, and the two foci of the coordinate system are located at $z = \pm a$.

\TwoPunctures~\cite{Ansorg:2004ds} adopts a PS-like coordinate system,
which is written in terms of coordinate variables $A\in[0,1)$,
$B\in[-1,1]$, and $\varphi\in[0,2\pi)$.
\TwoPunctures coordinates are related to Cartesian via\footnote{We
  swap the $x$ and $z$ coordinates of~\cite{Ansorg:2004ds} to simplify
  comparison.}
\begin{align}
  x &= b\frac{2A}{1-A^{2}}\frac{1-B^{2}}{1+B^{2}}\sin\varphi\,,\nonumber\\
  y &= b\frac{2A}{1-A^{2}}\frac{1-B^{2}}{1+B^{2}}\cos\varphi\,,\label{eq:twopuncturepaper_xyz}\\
  z &= b\frac{A^{2}+1}{A^{2}-1}\frac{2B}{1+B^{2}}\,.\nonumber
\end{align}
The two foci of \TwoPunctures coordinates are situated at \mbox{$z=\pm
  b$}. Of note, the coordinate $A$ is compactified with
\mbox{$|x|,|y|,|z|\to\infty$} as \mbox{$A\to1$}. Similar to the
term $\sin \nu$ (where $\nu\in[0,\pi)$) in PS coordinates, the
\mbox{$(1-B^{2})/(1+B^{2})$} (where $B\in[-1,1]$) term is a concave
down curve with a maximum of $1$ at the midpoint of the range of $B$
and zeroes at the endpoints $B=\pm1$. Unlike PS coordinates, however, the
coordinate system is not periodic in the variable $B$.

\nrpyell adopts the \nrpy PS-like coordinate system \SinhSymTP
$\left(\xxzero,\xxone,\xxtwo\right)$ with $\xxzero\in[0,1]$,
$\xxone\in[0,\pi]$, and $\xxtwo\in[-\pi,\pi]$. These are related to PS coordinates
via
\begin{align}
  a\sinh\mu &= \tilde{r}\equiv
  \sinhA\frac{\sinh(\xxzero/\sinhW)}{\sinh(1/\sinhW)}\,,\nonumber\\
  \nu &= \xxone\,,\label{eq:PSC_SinhSymTP}\\
  \varphi &= \xxtwo\,,\nonumber
\end{align}
with $\tilde{r}\in[0,\sinhA]$. Introducing the parameter $\bScale$, we
obtain
\begin{align}
  x &= \tilde{r}\sin(\xxone)\cos(\xxtwo)\,,\nonumber\\
  y &= \tilde{r}\sin(\xxone)\sin(\xxtwo)\,,\label{eq:SinhSymTP_xyz}\\
  z &= \left(\tilde{r}^{2} + \bScale^{2}\right)^{1/2}\cos(\xxone)\,.\nonumber
\end{align}
Given inputs for $\sinhW$, $\sinhA$ (domain size), and the foci parameter $\bScale$, \nrpy samples coordinates $\left(\xxzero,\xxone,\xxtwo\right)$ uniformly when setting up numerical grids. The foci exist at 
$z=\pm\bScale$, and grid point density near the foci can be increased or decreased by decreasing or increasing $\sinhW$, respectively. Contrast this to PS coordinates, in which there is no such parameter to adjust the focusing of gridpoints near foci. In fact when the foci separation is increased in PS coordinates, the density of gridpoints decreases in proportion.

\begin{figure}[!h]
  \centering
  \includegraphics[width=0.7\columnwidth]{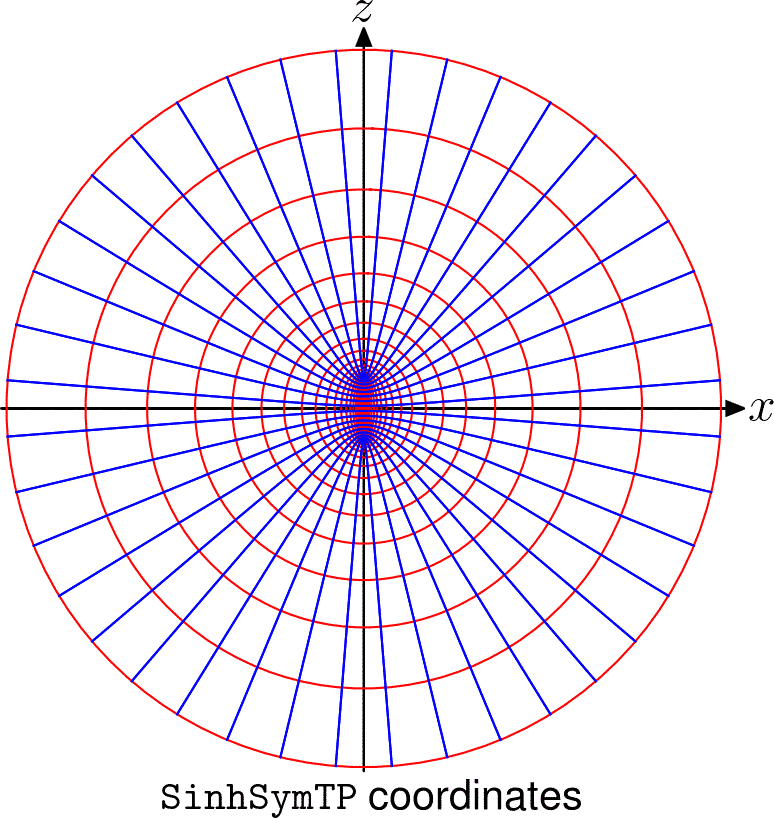}
  \caption{Curves of constant $\xxzero$ (red) and constant $\xxone$
    (blue) using a cell-centered grid structure with $\xxtwo=0$.}
  \label{fig:SinhSymTP}
\end{figure}

As illustrated in Fig.~\ref{fig:SinhSymTP}, like PS and \TwoPunctures coordinates, \SinhSymTP coordinates
become spherical in the region far from the foci. Note
also that regular spherical coordinates (with a non-uniform radial
coordinate) are fully recovered by setting $\bScale=0$.
As with \TwoPunctures, \nrpyellet possesses the option to rotate the
coordinate system, to situate the punctures on either the
$x$-axis ($x=\pm \bScale$) or the $z$-axis ($z=\pm \bScale$).

For all cases considered here, the outer boundary is set
to $10^6$ (i.e., $\sinhA = 10^6$), and the grid point spacing
parameter $\sinhW$ is set to $0.07$. We set $b$ so that the foci
match the punctures' positions and, when interpolating the ID to the
\etk grids, we adjust the origin of the coordinate system to coincide
with the center of mass of the punctures, as is conventional.

\subsection{Wavespeed}
\label{sec:wavespeed}
To propagate the hyperbolic system of equations forward in (pseudo)time from
the chosen initial conditions, we make use of \nrpy's method of lines
implementation. Specifically we choose the explicit fourth-order
Runge-Kutta (RK4) method. As this is an explicit method, and we use three
dimensions in space, the steps in time $\dt$ are constrained by the
CFL inequality:
\begin{equation}
  \label{eq:cfl_1d}
  \frac{c\dt}{\ds_{\rm min}} \leq \mathcal{C}_0 \,,
\end{equation}
where $c$ is the local wavespeed, and $\mathcal{C}_0$ is the CFL
factor, which depends on the explicit time stepping method and the
dimensionality of the problem. Empirically we find that
$\mathcal{C}_0 = 0.7$ ensures both stability and large time steps for both 2D and 3D cases presented in this paper, with one exception: when the 3D case is pushed to very high resolution. In the single highest-resolution 3D case in this work, we find $\mathcal{C}_0$ must be lowered to $0.55$ for stability.

Further, $\ds_{\rm min}$ is the minimum proper distance between
neighboring points in our curvilinear coordinate system:
\begin{equation}
  \ds_{\rm min} = \min \left(h_1 \dxxzero,\,h_2 \dxxone,\,h_3 \dxxtwo\right) \,,\label{eq:ds_i}
\end{equation}
where $h_i$ and $\dxxi$ are the $i$-th scale factor and grid spacing of the flat space metric, respectively.\footnote{In orthogonal, curvilinear coordinate systems, such as the ones supported by \nrpy, we have $h_{i}=\sqrt{\hat\gamma_{ii}}$ (no summation implied).}

The global relaxation time step $\dt_{\rm glob}$ is given by the minimum
value of $\ds_{\rm min}$ on our numerical grid. As
we adopt prolate spheroidal-like coordinates, the global
$\ds_{\rm min}$ occurs precisely at the foci of the coordinate
system. At this point, for simplicity we set the wavespeed $c=1$. As
this is merely a relaxation (as opposed to a physical) wavespeed, we
increase $c$ in proportion to the local $\ds_{\rm min}$, which
grows exponentially away from the foci. In this way we maintain
satisfaction of the CFL inequality while greatly improving the
performance of the relaxation method. Details and implications of our
implementation are described in Appendix~\ref{sec:appendix_wavespeed}.

\subsection{Choice of damping parameter $\eta$}
Sweeping through values of $\eta$ in early tests of our code (using different grid parameters than those adopted here), we found that setting $\eta = 12.5/M$ was optimal for minimizing the time to convergence of the relaxation scheme. After completing the tests, we later found that this is value of $\eta$ is not optimal for the grids and resolutions we chosen in this work. As a result, our core benchmark in Sec.~\ref{sec:3D_ID} below, comparing the performance of \nrpyell with \TwoPunctures is slower by roughly a factor of 2 as compared to the optimum $\eta$ for that grid choice, which would be closer to $\eta=18/M$. We note that this sub-optimal choice does not influence our relaxed solution, as every choice of $\eta$ that leads to a stable evolution yields that same relaxed solution.

To find the optimal value of $\eta$, we perform a linear sweep through possible values in a range $\eta=1$ through 20, in intervals of $0.25$ (recall $\eta$ has units of inverse mass, $1/M$, and here when quoting values of $\eta$ we always choose $M=1$).
In the steady-state regime, the $L^2$-norm of the relative error, $E$ (see Sec.~\ref{sec:results} for
details), becomes constant. Therefore once we have chosen a trial value of $\eta$, we perform the relaxation until $|dE/dt| < \delta$ for $n$ consecutive iterations, where $\delta \sim 10^{-4}{-}10^{-6}$ and $n\sim10{-}100$. I.e., we relax until the $L^2$-norm of the relative error has reached its minimum, constant value. The optimum value of $\eta$ is that value that results in the fewest relaxation iterations before the $L^2$-norm of the relative error has plateaued.

As $\eta$ has units of $1/t$, it is natural to inquire whether the optimum damping parameter,
$\eta_{\rm optimum}$, is inversely proportional to the relaxation-wave-crossing time,
$T_{\rm RC}$. We applied the aforementioned optimum $\eta$ search to a variety of grids, each with its own $T_{\rm RC}$ (in the range $T_{\rm RC} \in (0.24,0.99)$), finding that the optimum $\eta$ obeys the following power law:\footnote{For this fit, we generated 3-dimensional initial data with physical parameters as described in Sec.~\ref{sec:3D_ID} (with $M=1$ at resolutions ranging from $32^2{\times}16$ to $128^2{\times}16$). As the wavespeed depends on grid spacing, the relaxation-light-crossing time is affected by the resolution.}
\begin{equation}
\label{eqn:eta_power_law}
    M \eta_{\rm optimum} \approx \frac{6.9}{(T_{\rm RC}/M)^{0.79}} - 0.7 .
\end{equation}
%


\section{Results}
\label{sec:results}
Validation of \nrpyell is performed in two stages. First we
generate initial data (ID) for a given physical scenario with the
widely used \TwoPunctures~\cite{Ansorg:2004ds} code, increasing
resolution on the \TwoPunctures grids until roundoff error dominates
its numerical solution of \eqref{eqn:hamiltonian_constraint_v3},
$u$. We refer to this high-resolution result as the trusted
solution. Second we generate the same ID with \nrpyell, and
demonstrate that its results approach the trusted solution at the expected convergence rate.

We repeat this procedure twice: first for an axisymmetric
case of two equal-mass BHs with spin vectors collinear with their
separation vector, and second for a full 3D case involving a
GW150914-like unequal-mass, quasi-circular, spinning BBH system. To
demonstrate the fidelity of the latter case, we first generate
\nrpyell and \TwoPunctures ID at similar levels of accuracy. Then,
using the \etk~\cite{Loffler:2011ay,etk_web,Cactusweb} we
evolve the ID through inspiral, merger, and ringdown, and compare
the results. Finally we note
that $M=M_{+}+M_{-}$ is defined as
the sum of individual ADM masses of the punctures (Eq.~(83) of
Ref.~\cite{Ansorg:2004ds}).

\subsection{Axisymmetric initial data}
\label{sec:axi_ID}
\begin{figure}[!h]
	\begin{tabular}{c}
		\includegraphics[width=0.99\columnwidth,clip]{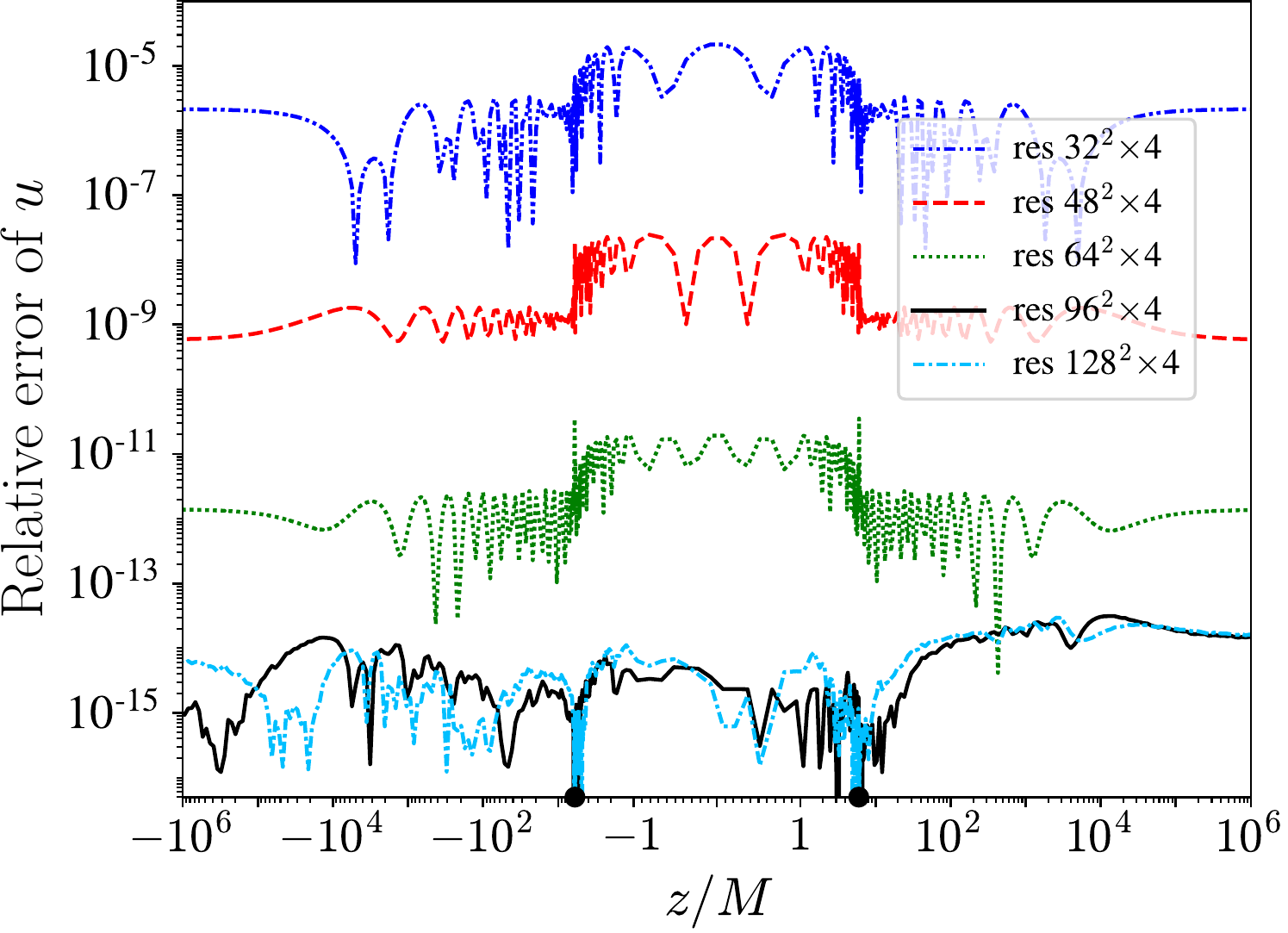}\\
		\includegraphics[width=0.99\columnwidth,clip]{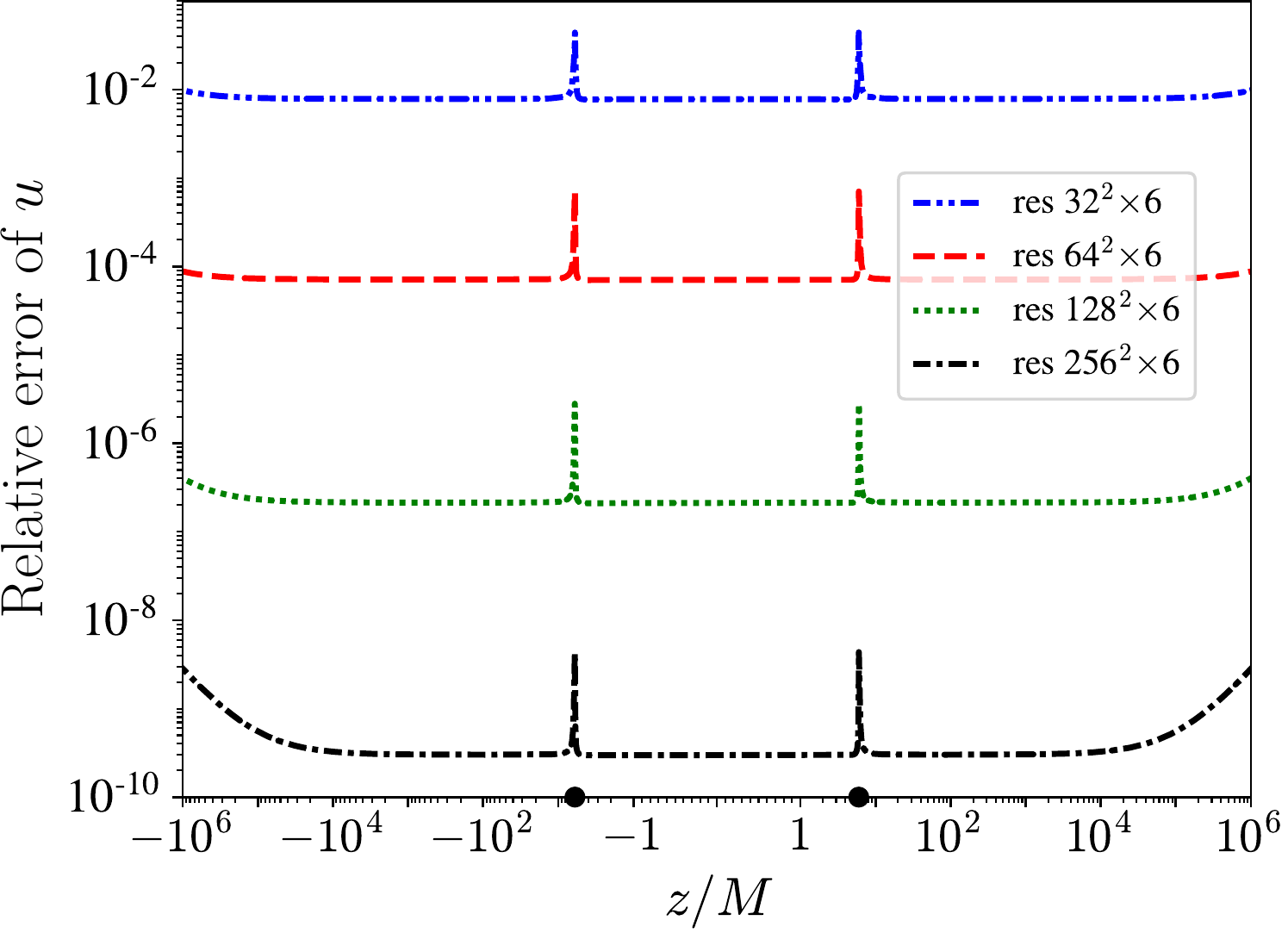}
	\end{tabular}
	\caption{Calibration of \TwoPunctures and \nrpyell solutions, axisymmetric ID study.
          \textbf{Top}: Relative errors between a super-high resolution
          ($196^2 {\times} 4$) and lower-resolution results from the \TwoPunctures code.
	  \textbf{Bottom}: Calibration of \nrpyell solutions at
          various resolutions ($\Nzero {\times} \None {\times} 6$),
          comparing against the trusted solution (i.e., the
          \TwoPunctures result at $96^2 {\times} 4$ resolution). The
          horizontal axis is logarithmic in the range $|z|>1$ and
          linear otherwise. Further, the black dots on the horizontal axes denote the puncture BH positions (also the locations of the coordinate foci). \TwoPunctures data were rotated so that the punctures and the foci lie on the $z$ axis.
	  \label{fig:axisymmetric_calibration}
	}
\end{figure}
In this test, we generate initial data (ID) for a scenario symmetric about
the $z$-axis: two equal-mass punctures at
rest, with spin components $\vec{S}_{\pm} = \pm
0.4M_{\pm}^{2}\hat{z}$ and initial positions $z_{\pm}=\pm6M$.\footnote{The
bare mass of each puncture is $m=0.456428$ and the total ADM mass is
$M_{\rm ADM}=0.979989M$.}

We start by constructing the trusted solution $u$, against which
we will compare results from \nrpyell. The trusted solution is
generated by increasing the resolution of the \TwoPunctures numerical
grids until roundoff errors dominate. As we adopt double-precision
arithmetic, this occurs when relative errors reach levels of roughly
$10^{-14}$. The top panel of Fig.~\ref{fig:axisymmetric_calibration}
indicates that this occurs at a \TwoPunctures grid resolution of
$N_A\times N_B\times N_{\phi}=96^2\times 4$.\footnote{To ensure we reach
  the level where roundoff errors dominate, the maximum number of
  iterations of the Newton-Raphson method (\texttt{Newton\_maxit}
  parameter) is set to $10$, and the tolerance
  (\texttt{Newton\_tol} parameter) is to $\times 10^{-16}$, for the
  axisymmetric case. Also we fix $N_{\phi}=4$, following an
  axisymmetric example in~Ref.~\cite{Ansorg:2004ds}.}

After obtaining this $96^2\times 4$ trusted solution, we next compare
  it against \nrpyell at various resolutions in the bottom panel of
  the figure. To allow for a point-wise comparison, we use the spectral interpolator in \TwoPunctures~\cite{Paschalidis:2013oya} to evaluate the \TwoPunctures solution at every point on our grid.
  Notice that the numerical errors drop by roughly $2^9$
  each time the resolution is doubled, indicating that numerical
  convergence is dominated by our choice of $10^{\rm th}$-order finite-difference (FD) stencils. That perfect $10^{\rm th}$-order convergence is
  not obtained is unsurprising, as our outer boundary
  condition is approximate and implements its own $6^{\rm th}$-order
  FD stencils.\footnote{\nrpy currently requires the minimum number of grid points in any
  given direction to be even and larger than $\mathtt{FD\_order}/2$,
  where $\mathtt{FD\_order}$ order of the finite difference
  scheme. Thus we set $\Ntwo = 6$ despite the system being
  axisymmetric. This requirement may be relaxed in the future.}

\begin{figure}
	\begin{tabular}{c}
		\includegraphics[width=0.99\columnwidth,clip]{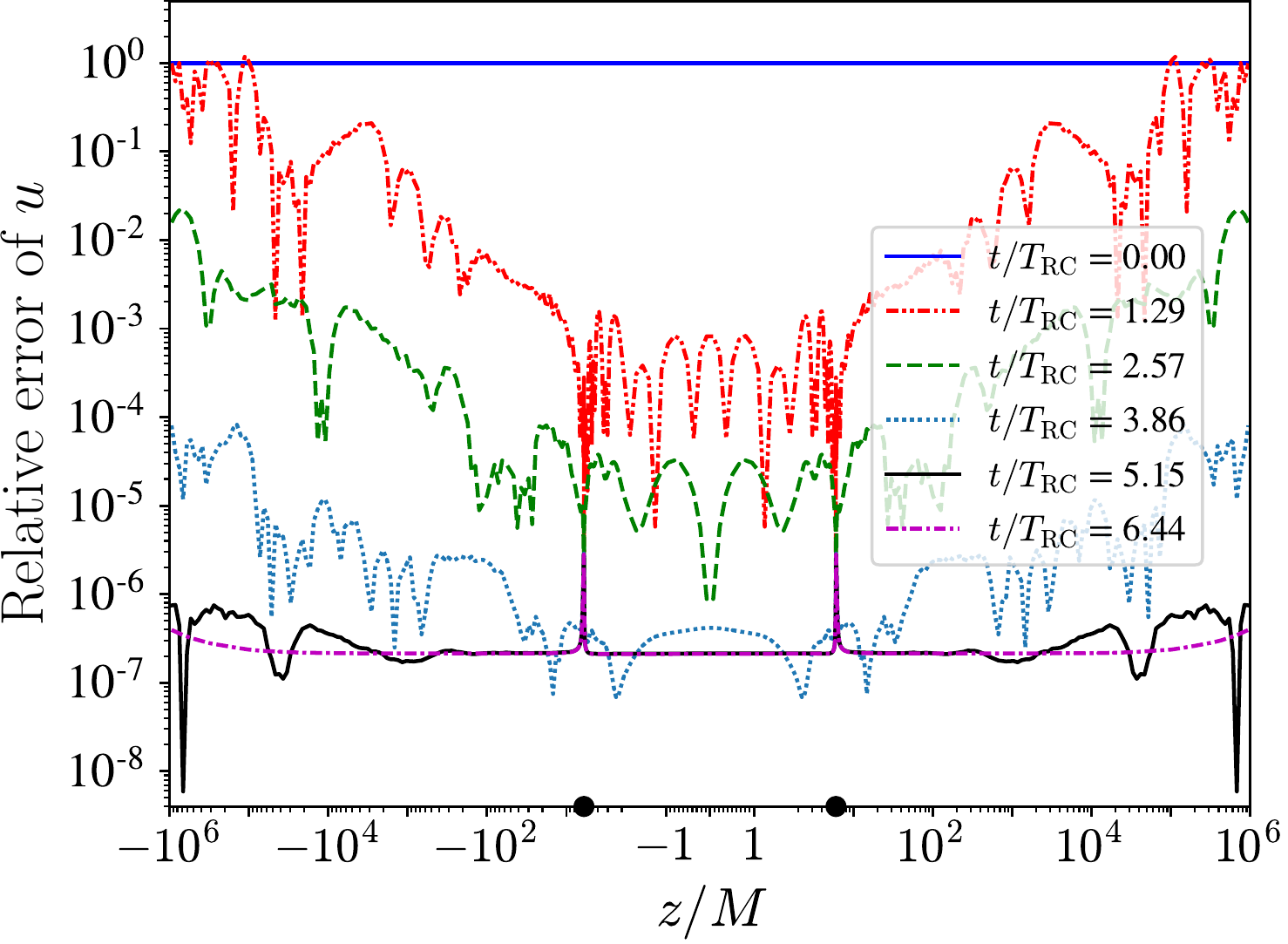}\\
		\includegraphics[width=0.99\columnwidth,clip]{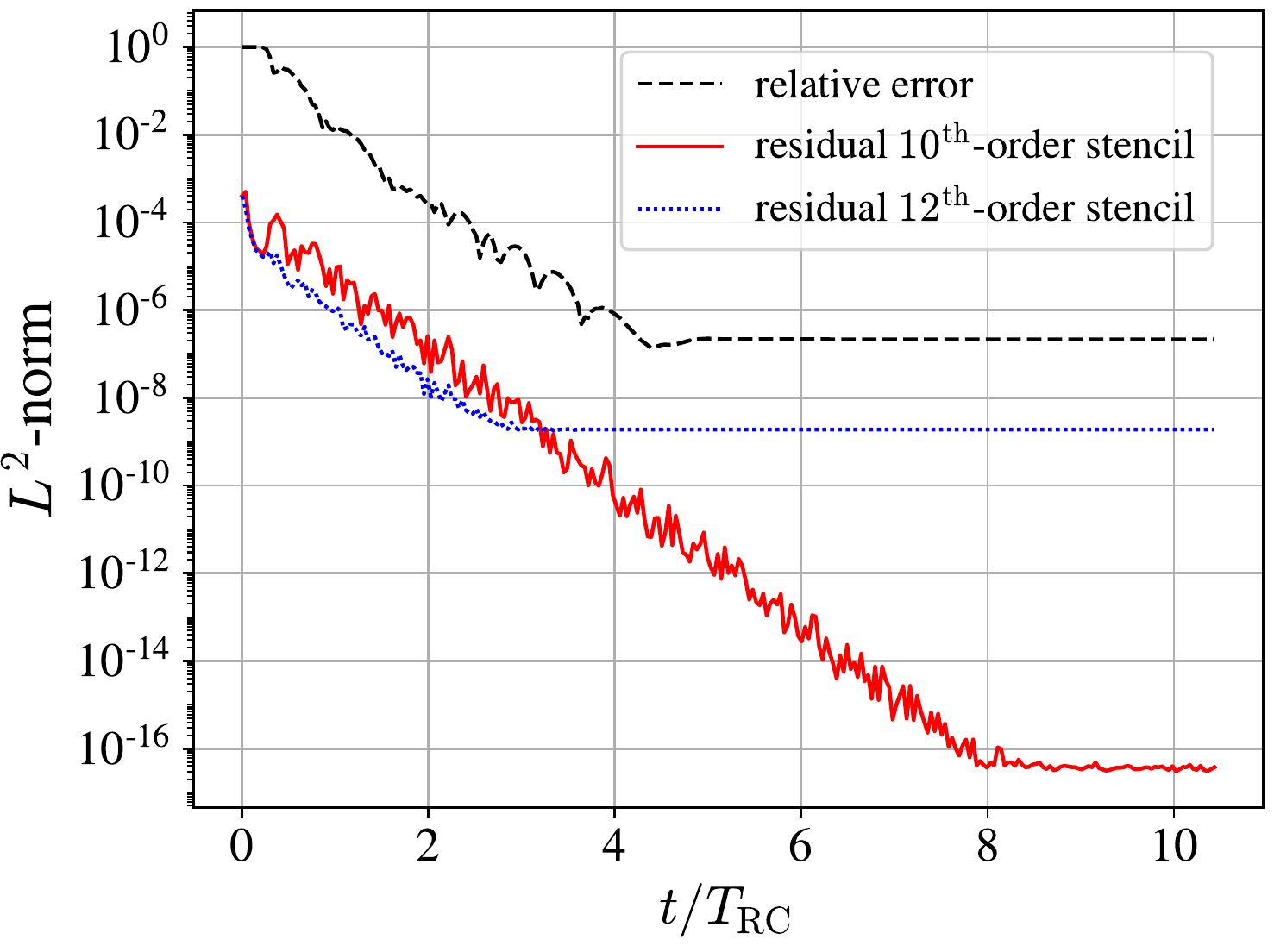}
	\end{tabular}
	\caption{Axisymmetric ID study.
		\textbf{Top}: Snapshots of the relative error between the
                \nrpyell solution at $128^2{\times 4}$ resolution and
                the trusted solution at different times. \TwoPunctures data were rotated so that the punctures and the foci lie on the $z$ axis.
		\textbf{Bottom}: Black dashed curve: $L^{2}$-norm of
                the same relative error as the top panel, computed
                within a sphere with radius $r=100M$ and as a function
                of time. Blue (dotted) and red (solid) curves: $L^{2}$-norms of residuals
                (second term in~\eqrefalt{eqn:hamiltonian_constraint_v3}) computed
                within the same sphere at two different
                finite-difference orders as a function of time.
		\label{fig:axisymmetric_relaxation}
	}
\end{figure}

The exponential convergence in time of the hyperbolic relaxation method
can be readily seen in Fig.~\ref{fig:axisymmetric_relaxation}. In the
top panel snapshots of the relative error at resolution
  $128^2{\times}6$ are shown as a function of relaxation-wave-crossing times, or RCTs (i.e., the time required for a relaxation wave to cross the numerical domain; see
  Appendix~\ref{sec:appendix_wavespeed}). During the first RCT, the errors decrease (increase) in
the region near (away from) the punctures and subsequently drop
exponentially in time. Numerical errors are consistently smaller
  in the region near the origin due to the extreme grid focusing
  there. Notice that after ${\sim}4$
RCTs the relative errors between the punctures have already reached the same order of magnitude as that
of the fully relaxed solution, and the remaining relaxation time (40\%
of the total) is spent decreasing the relative errors
away from the punctures.

Coincident with the exponential relaxation along the $z$-axis,
  the $L^2$-norm of both the relative errors and residual (second term in the left-hand side of
\eqrefalt{eqn:hamiltonian_constraint_v3}), computed inside a sphere of
  radius $100M$ around the origin (situated at the center of mass), displays a steady exponential drop,
  as illustrated in the bottom panel of Fig.~\ref{fig:axisymmetric_relaxation}.
We find that when the FD  order for computing the residual matches the one used by the relaxation
algorithm, the $L^{2}$-norm of the residual continues to
drop even after the steady state has been reached, until roundoff errors dominate this measure of residual. This indicates that the residual has become
overspecialized to the approximation adopted for the FD operators.
To ameliorate this, we also compute the residual using 12\textsuperscript{th}-order stencils,
finding this measure of the residual to have more consistent behavior
with the true error (i.e., the relative error against the trusted solution). We repeated this
analysis at 8\textsuperscript{th} and 14\textsuperscript{th} finite-differencing order and
found the same qualitative behavior as 12\textsuperscript{th} order.

\subsection{Full 3D initial data}
\label{sec:3D_ID}

The second test consists of two punctures with mass ratio \mbox{$q = 36/29$}, in a quasicircular orbit emulating the GW150914 event~\cite{Abbott:2016blz}. The punctures are located at \mbox{$z = \pm 5M$}, with spin components \mbox{$\vec{S}_{+} = 0.31 M_{+}^2 \hat{y}$} and  \mbox{$\vec{S}_{-} = -0.46 M_{-}^2 \hat{y}$}.%
\footnote{The bare (local ADM) masses of the punctures are \mbox{$m_{+} = 0.518419$} (\mbox{$M_{+} = 0.553846$}) and \mbox{$m_{-} = 0.391936$.} (\mbox{$M_{-} =0.446154$}). The total ADM mass is $M_{\rm{ADM}} = 0.989946M$. To elicit a quasicircular orbit, linear momenta are set to \mbox{$\vec{P}_{\pm} = \pm P_{\phi} \hat{x}  \pm P_r \hat{z}$}, where \mbox{$P_{\phi} = 9.53{\times}10^{-2}$} and \mbox{$P_r =-8.45{\times}10^{-4}$, in code units.}%
}

\begin{figure}[!h]
	\begin{tabular}{c}
		\includegraphics[width=0.99\columnwidth,clip]{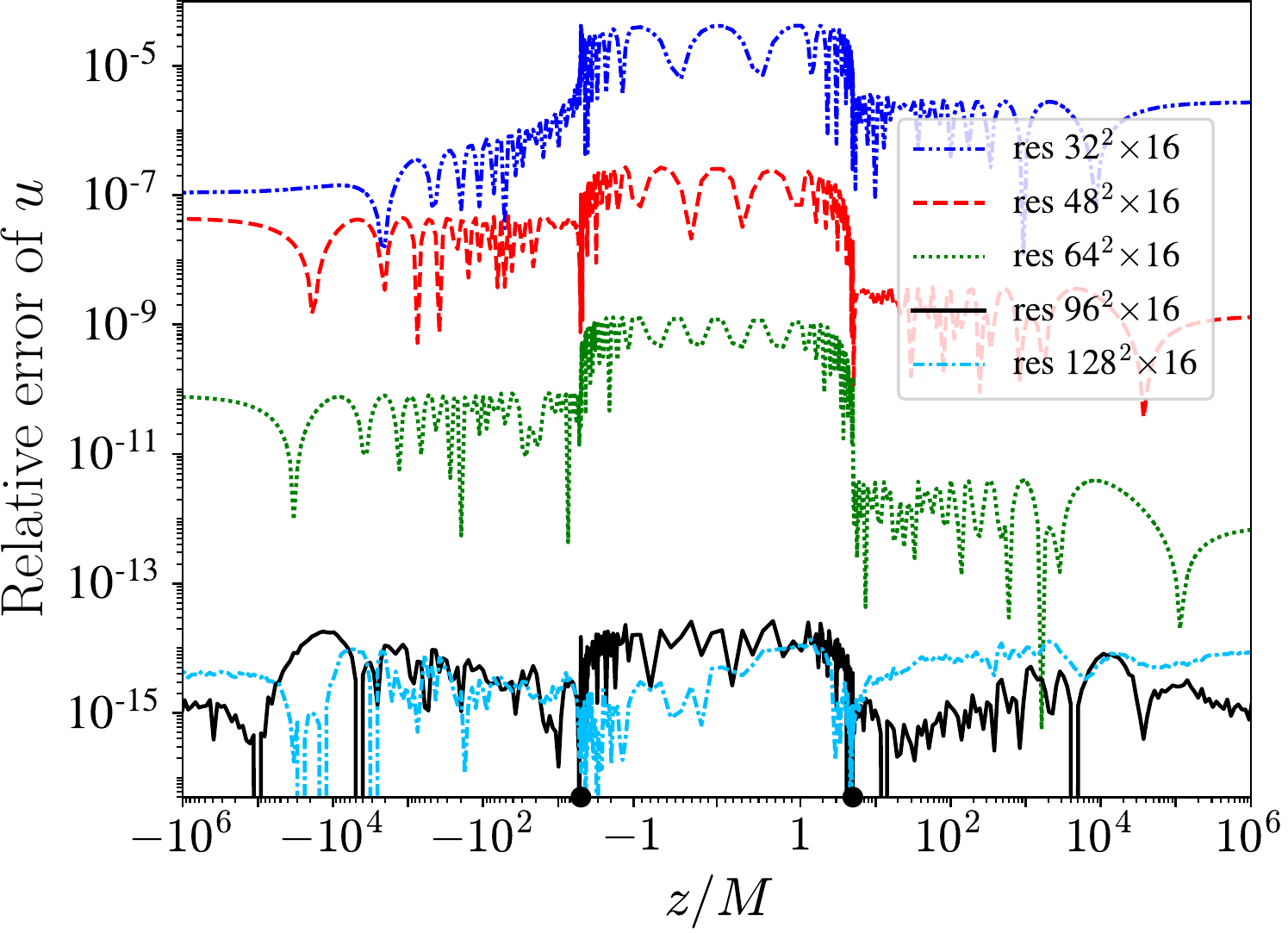}\\
		\includegraphics[width=0.99\columnwidth,clip]{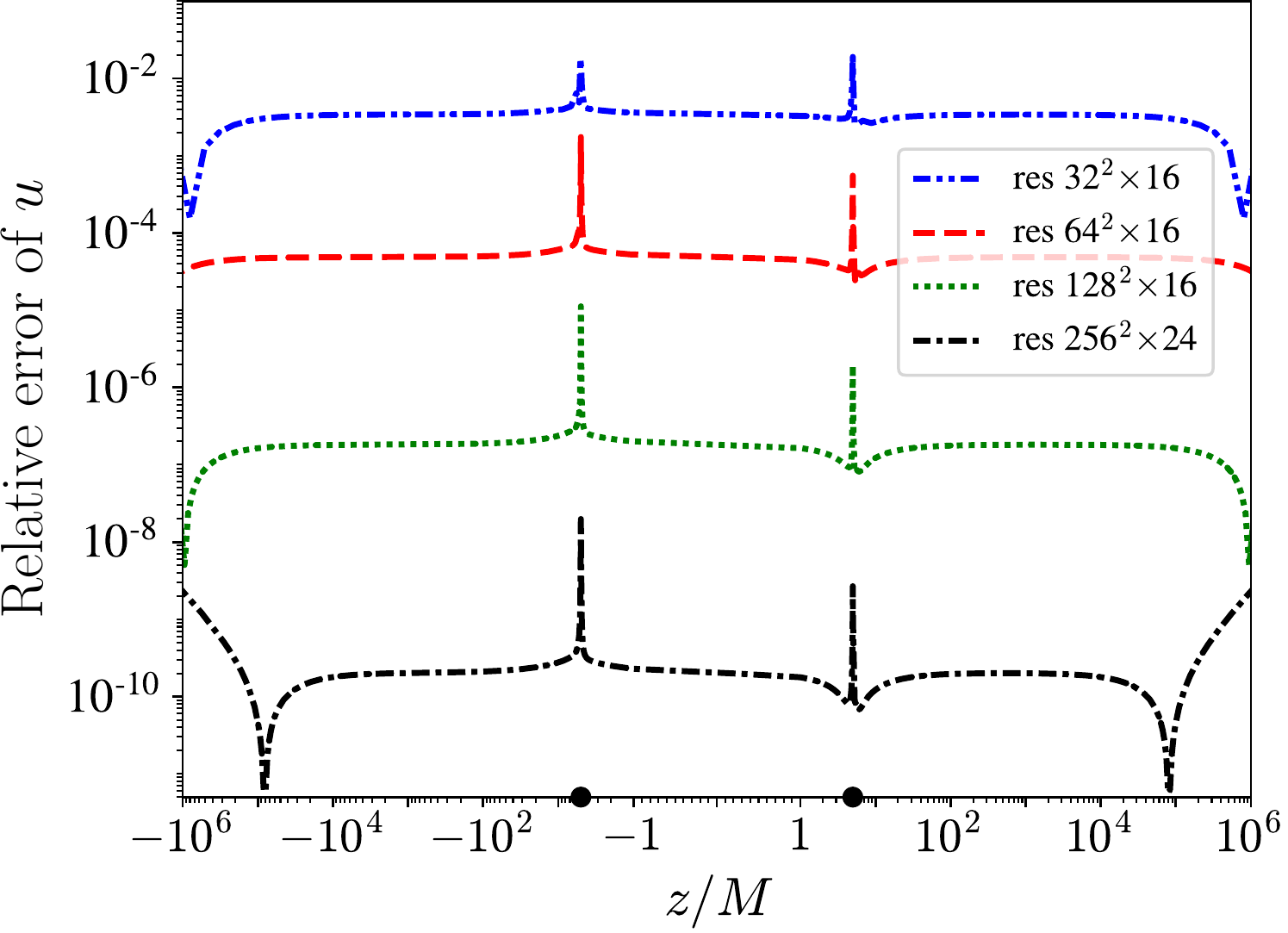}
	\end{tabular}
	\caption{Same as Fig.~\ref{fig:axisymmetric_calibration}, but
          for GW150914-like, full-3D BBH initial data.
		\label{fig:quasicircular_calibration}
	}
\end{figure}

The validation procedure for both \TwoPunctures and \nrpyell follows the prescription used in the axisymmetric case. Since the full-3D system does not posses axial symmetry, the number
of grid points along the azimuthal direction is no longer minimal. As indicated in the top
panel of Fig.~\ref{fig:quasicircular_calibration}, the trusted \TwoPunctures solution has
resolution $96^2{\times}16$, and we found that using $N_{\phi} > 16$
does not lower the relative errors further. In \nrpyell (bottom panel of Fig.~\ref{fig:quasicircular_calibration}), the optimal resolution in the azimuthal direction was also found to be $\Ntwo = 16$, except for when the highest resolution ($256^2\times 24$) was chosen. We find this solution to have smaller
relative errors when compared to the $256^2\times 16$ resolution data (not shown), and
requires a lower CFL-factor of $\mathcal{C}_0 = 0.55$ for numerical stability. With
truncation errors dominated by sampling in the $\xxzero$ and $\xxone$ directions, we find roughly $9^{\rm th}$-order convergence for this case as well.

\begin{figure}[!h]
	\begin{tabular}{c}
		\includegraphics[width=0.99\columnwidth,clip]{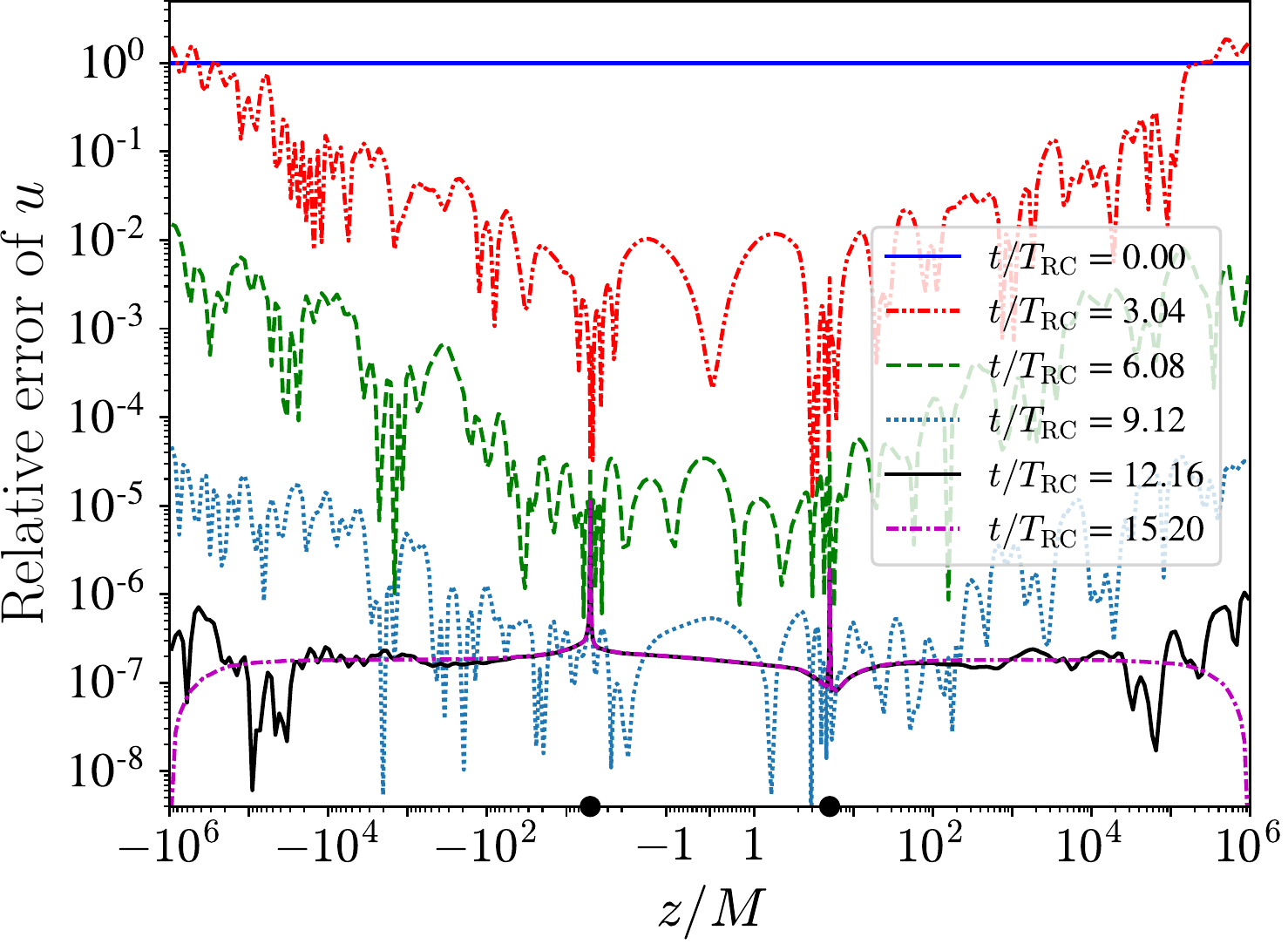}\\
		\includegraphics[width=0.99\columnwidth,clip]{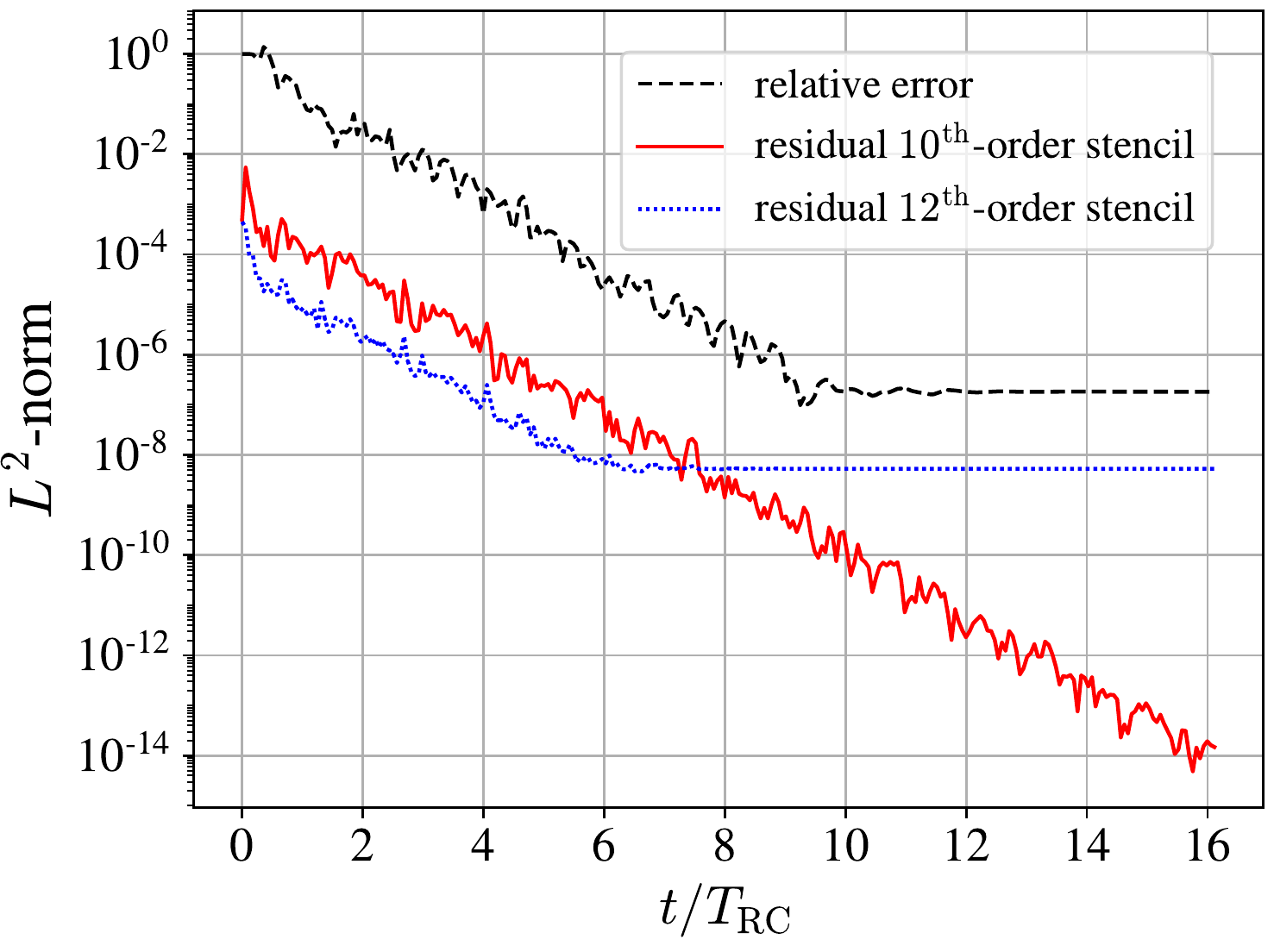}
	\end{tabular}
	\caption{Same as Fig.~\ref{fig:axisymmetric_relaxation}, but
          for GW150914-like, full-3D BBH initial data, with \nrpyell resolution of $128^2 {\times} 16$.
		\label{fig:quasicircular_relaxation}
	}
\end{figure}

As expected, the relaxation to the steady state has exponential convergence in (pseudo)time
as depicted in both panels of Fig.~\ref{fig:quasicircular_relaxation}. Although the
relaxation requires a larger number of RCTs in the full-3D case, its qualitative behavior is
the same.

\begin{figure}[!h]
	\centering
	\includegraphics[width=\columnwidth,clip]{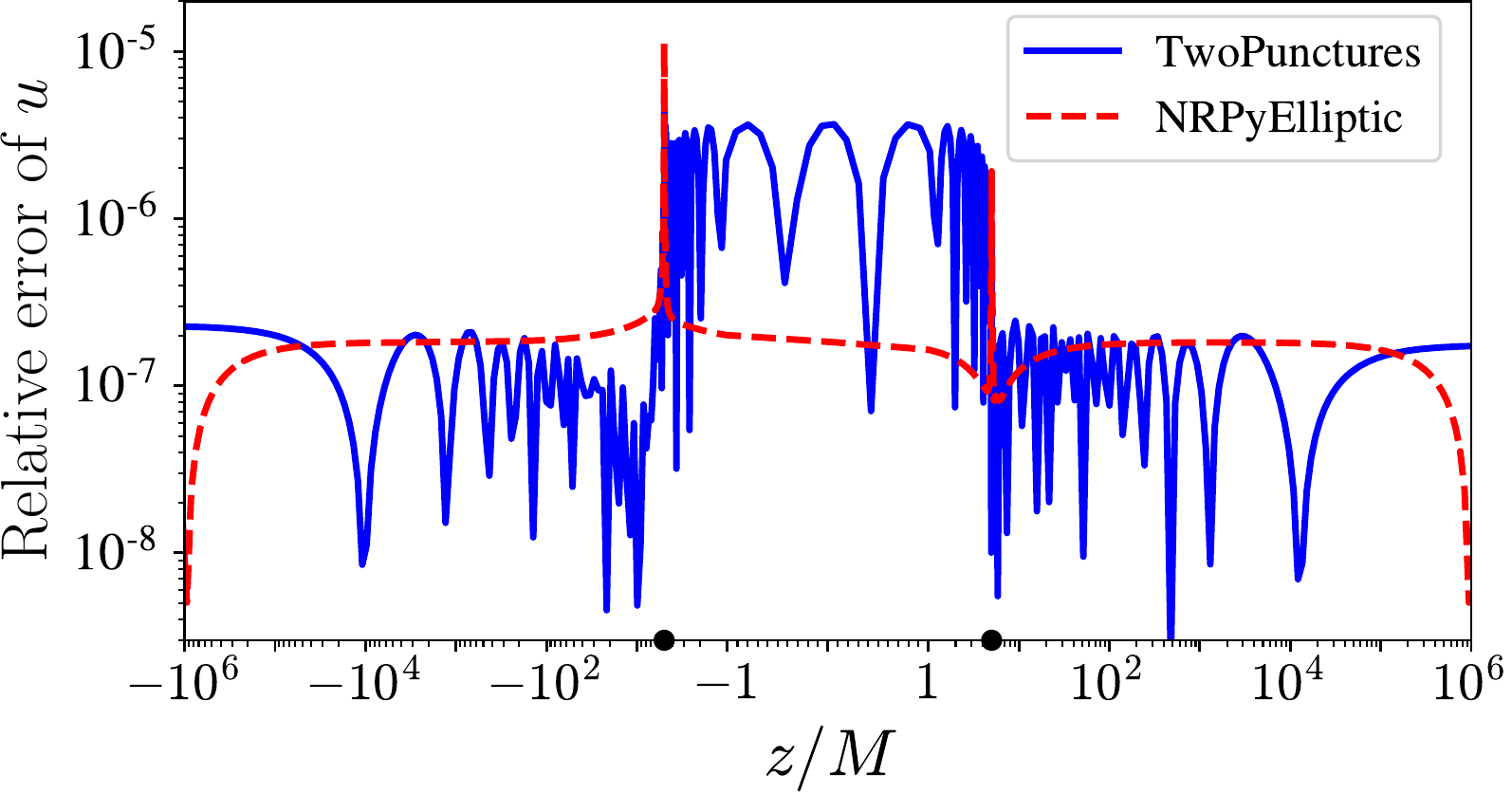}
	\caption{Initial data quality of the GW150914-like BBH 3D scenario evolved with the \etk, showing the relative error of $u$ against the trusted \TwoPunctures solution. Here \TwoPunctures and \nrpyell ID used resolutions of $38^2\times 16$ and $128^2\times 16$
          respectively. Note that \TwoPunctures data were rotated so that the punctures are positioned on the $z$ axis.}
	\label{fig:TP_vs_NE_uu}
\end{figure}

Next, to demonstrate the fidelity of full-3D initial data generated by
\nrpyell, we first generate ID for the GW150914-like BBH scenario
using both \nrpyell and \TwoPunctures. These ID are chosen to be of
comparable quality, as shown in Fig.~\ref{fig:TP_vs_NE_uu}. Precise numerical
parameters for the ID exactly follow the GW150914 \etk
gallery example~\cite{wardell_barry_2016_155394,
	Loffler:2011ay,
	Pollney:2009yz,
	Schnetter:2003rb,
	Thornburg:2003sf,
	Ansorg:2004ds,
	Dreyer:2002mx,
	Goodale02a,
	Brown:2008sb,
	Husa:2004ip,
	Thomas:2010aa}, for which complete results are freely available online.\footnote{See
	\url{https://einsteintoolkit.org/gallery/bbh/index.html} for more
	details.} Notably
the \TwoPunctures code required slightly less than a minute to
generate the ID (we input the bare masses directly), while \nrpyell
required about 10 minutes, for a difference in performance of
${\approx}12$x.\footnote{Benchmarks were performed on a 16-core AMD Ryzen 9 3950X CPU; the speed-up factor was found to be very similar on other CPUs as well. Moreover, the damping parameter was not set to the optimum value of $\eta = 18/M$, and if it were the difference in performance would drop to only $\approx$6x.}

\begin{figure*}[!htb]
	\includegraphics[width=0.8\textwidth,clip]{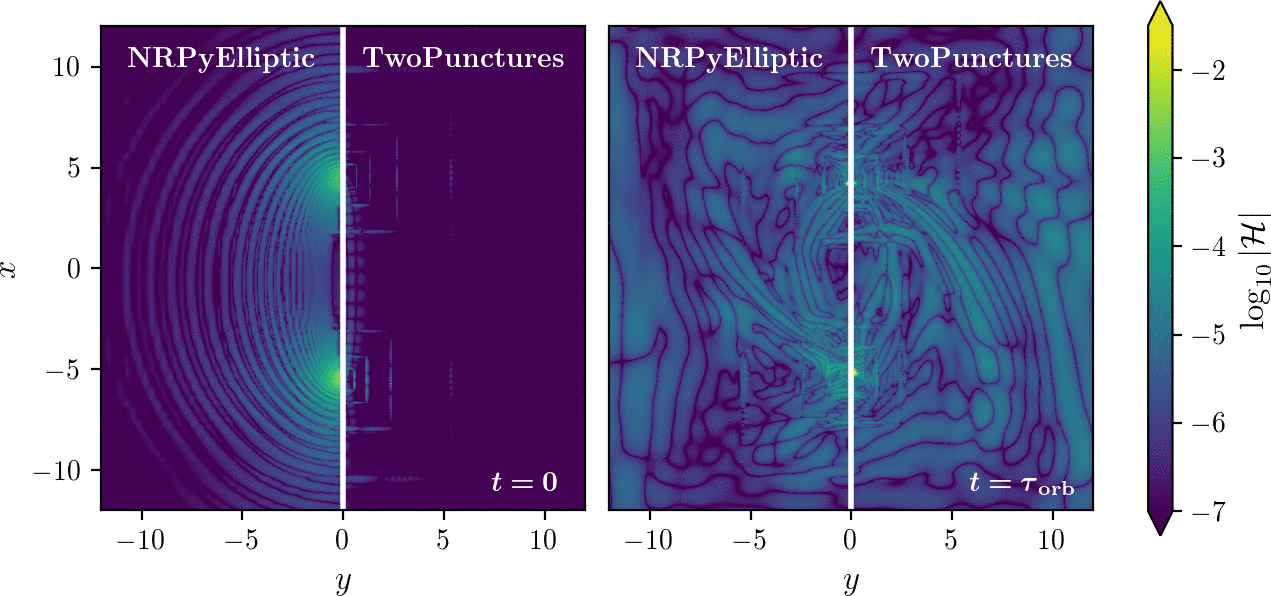}
	\caption{Hamiltonian constraint violation
		$\log_{10}\left|\mathcal{H}\right|$, as defined in~\eqref{eqn:hamiltonian_constraint}, in orbital plane after
		interpolation to AMR grids at $t=0$ (\textbf{left}) and after one
		orbit (\textbf{right}). Note that \nrpyell data were rotated so that the punctures are positioned on the $x$ axis.}
	\label{fig:TP_vs_NE_constraint_violations}
\end{figure*}

After generation, the ID from both codes are interpolated onto the
evolution grids, which consist of a set of
\texttt{Carpet}~\cite{Schnetter:2003rb}-managed Cartesian AMR patches
in the strong field region, with resolution
around the punctures of ${\approx}M/52$. These AMR patches are
surrounded by a \texttt{Llama}~\cite{Thornburg:2003sf}-managed cubed
spheres grid in the weak-field region. When interpolating the ID onto
these grids, \TwoPunctures adopts a high-accuracy spectral
interpolator (see e.g., \cite{Paschalidis:2013oya}), and \nrpyell uses
the third-order Hermite polynomial interpolator from the
\texttt{AEILocalInterp} thorn. As our choice of Hermite interpolation order results in far less accuracy, it introduces error to the Hamiltonian
constraint violation at time $t=0$ on the orbital plane, as shown
in the left panel of Fig.~\ref{fig:TP_vs_NE_constraint_violations}.

We then evolve these ID forward in time using the \texttt{McLachlan}
BSSN thorn~\cite{Brown:2008sb}. Notice that after the first orbital period
the Hamiltonian constraint violations on the orbital plane are essentially
identical (right panel of
Fig.~\ref{fig:TP_vs_NE_constraint_violations}), indicating that
numerical errors associated with the evolution quickly dominate ID errors.
\begin{figure}[!h]
	\centering
	\includegraphics[width=\columnwidth,clip]{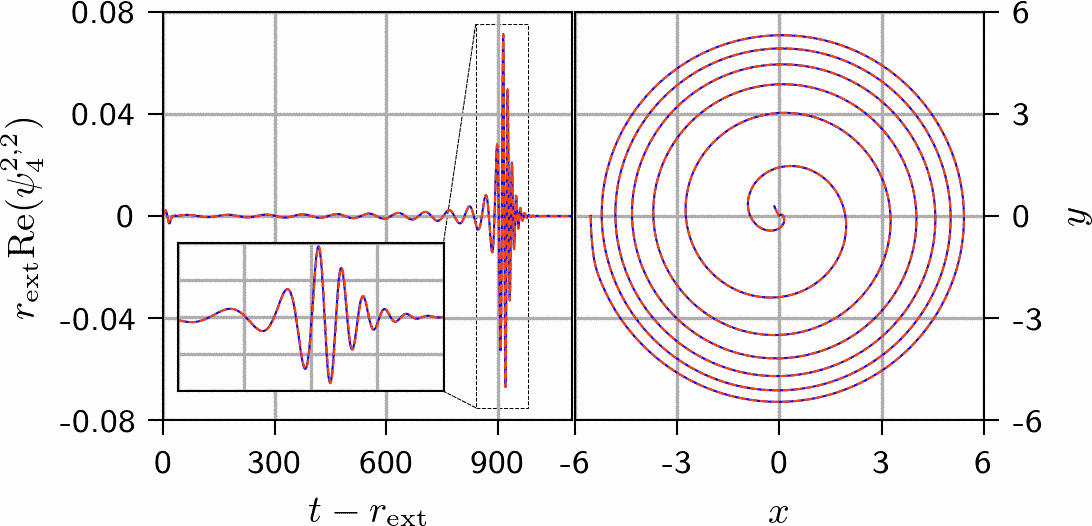}
	\caption{Simulation results with ID by \TwoPunctures (solid, blue) and \nrpyell (dashed, orange-red). \textbf{Left}: Dominant mode ($\ell=2$, $m=2$) of the Weyl
		scalar $\psi_{4}$ at extraction radius $r_{\rm
                  ext}=500M$. \textbf{Right}: Trajectory of the less-massive BH.}
	\label{fig:TP_vs_NE_psi4_and_trajectories}
\end{figure}

Finally, to demonstrate that \nrpyell's 3D data are of sufficiently high fidelity, we demonstrate that when they are evolved forward in time the results are indistinguishable from an evolution of \TwoPunctures ID at comparable initial accuracy. For instance in
Fig.~\ref{fig:TP_vs_NE_psi4_and_trajectories} we show the
trajectory of the less-massive BH and the dominant mode ($\ell=2$,
$m=2$) of the Weyl scalar $\psi_{4}$, for both evolutions. We find that quantities extracted from the evolution of \nrpyell ID to be in excellent agreement with
results from evolution of ID from the trusted \TwoPunctures code. Although not shown, the trajectory of the
more-massive BH for both simulations are also visually
indistinguishable, as are other, higher-order $\psi_{4}$ modes.


\section{Conclusions and Future Work}
\label{sec:conclusions}
\nrpyell is an extensible ID code for
numerical relativity that recasts non-linear
elliptic PDEs as covariant, hyperbolic PDEs. To this end it adopts
the hyperbolic relaxation method of~\cite{Ruter:2017iph}, in
which the (pseudo)time evolution of the hyperbolic PDEs
exponentially relaxes to a steady state consistent with the solution
to the elliptic problem. That standard hyperbolic methods are used to
solve the elliptic problem is beneficial, as consumers of
numerical relativity ID generally already have expertise in
solving hyperbolic PDEs. Thus the learning curve is significantly
lowered for core users of \nrpyell, enabling them to build on
existing expertise to modify and extend the solver.

\nrpyell leverages \nrpy's reference metric infrastructure to solve the hyperbolic/elliptic PDEs in a wide variety of
Cartesian-like, spherical-like, cylindrical-like, and bispherical-like coordinate systems. As its first
application, in \nrpyell we solve a nonlinear elliptic PDE to set up
two-puncture ID for numerical
relativity. Similar to the \TwoPunctures~\cite{Ansorg:2004ds} code,
\nrpyell solves the elliptic PDE for the Hamiltonian constraint in a prolate spheroidal-like
coordinate system. But unlike the one-parameter coordinate system used
in prolate-spheroidal coordinates or \TwoPunctures coordinates, \nrpyell adopts \nrpy's \SinhSymTP three-parameter
coordinate system, providing greater flexibility in setting up the
numerical grid.

As the \SinhSymTP numerical grid is not compactified, finite-radius
boundary conditions must be applied. To address this, a new radiation
boundary condition algorithm within \nrpy has been developed, which is
based on the widely used \newrad radiation boundary condition driver
within the \etk. While \newrad implements a second-order approach,
\nrpyell extends to fourth and sixth finite difference orders as
well. This high-order boundary condition meshes well with
the high (tenth) order finite-difference representation of the elliptic
operators adopted in \nrpyell.

To greatly accelerate the relaxation, we set the wavespeed of the
hyperbolic PDEs to grow in proportion to the grid spacing. As the
\SinhSymTP grid spacing grows exponentially with distance from the
central region of the coordinate system, so does the relaxation
wavespeed. Thus this approach is many orders of magnitude faster than
the traditional, constant-wavespeed choice, in fact making it fast
enough to set up high-quality, full-3D BBH ID for numerical
relativity.

Although \nrpyell currently requires only a tiny fraction of the
total runtime of a typical NR BBH merger calculation, it is roughly 12x slower than \TwoPunctures when setting
up the full-3D BBH ID in this work. Efforts in
the immediate future will in part focus on improving this
performance. To this end, a couple of ideas come to mind. First, all ID generated in this work used
the trivial ($u=v=0$) initial guess. We plan to explore whether
relaxations at lower-resolutions might be used to provide a superior
initial guess on finer grids, in which convergence is accelerated.

Second, due to the CFL condition, the speed of \nrpyell is
proportional to the smallest grid spacing, which occurs at the foci of
our \SinhSymTP coordinate system. Typical grid spacings at the foci
are ${\sim}10^{-4}M$ due to extreme grid focusing there, which
in turn are ${\sim}1/100$ those typically used in (near-equal-mass) binary
puncture evolutions, indicating that a significant speed-up may be
possible if superior grid structures are used.

\begin{figure}[!h]
  \includegraphics[width=0.8\columnwidth]{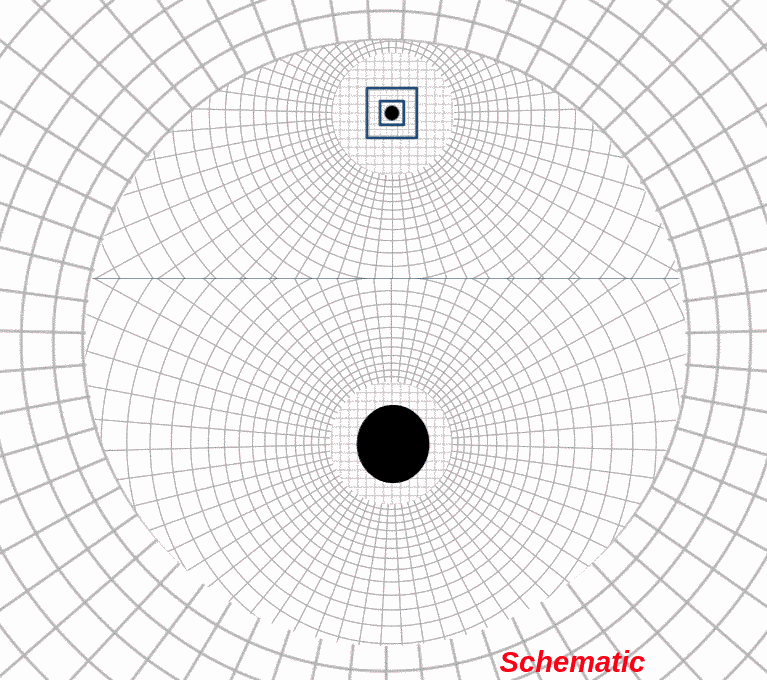}
  \caption{Schematic of seven-grid BiSpheres numerical meshes used for
    \bhah ${\sim}$6:1 mass-ratio BBH simulations during the long
    inspiral phase. BHs are represented as black dots.}
  \label{fig:bhah_grid}
\end{figure}

To this end, we plan to adopt the same seven-grid
bispheres grids infrastructure adopted by the \nrpy-based
\bhah~\cite{bhah_web} project. As illustrated
in~Fig.~\ref{fig:bhah_grid}, seven-grid BiSpheres consists of seven
overlapping spherical-like and Cartesian-like grids. This approach
places Cartesian-like AMR grid patches over regions where the
spherical-like grids would otherwise experience extreme grid focusing
($r\to 0$), constraining the smallest grid spacings in the
strong-field region to ${\approx}M/200$. Thus with such grids, accounting for needed inter-grid interpolations, we might
expect roughly a ${\sim}10$x increase in speed---making \nrpyell comparable in performance to
\TwoPunctures for near-equal-mass-ratio systems---all while
maintaining excellent resolution in the strong-field region. Further,
the Cartesian-like AMR patches on these grids are centered precisely at
the locations of the compact objects, making them efficiently
tunable to higher mass ratios, unlike \SinhSymTP or other
prolate spheroidal-like coordinates mentioned in this work. Extending
\nrpyell to higher mass ratios in this way, as well as to other types
of NR ID, will be explored in forthcoming papers.

\section*{Acknowledgments}
The authors would like to thank S.~Brandt and R.~Haas
for useful discussions and suggestions during the preparation of
\nrpyell. Support for this work was provided by NSF awards
OAC-2004311, PHY-1806596, PHY-2110352, as well as NASA awards ISFM-80NSSC18K0538
and TCAN-80NSSC18K1488. Computational resources were in part provided by
West Virginia University's Thorny Flat high-performance computing
cluster, funded by NSF MRI Award 1726534 and West Virginia
University. T.P.J. acknowledges support from West Virginia University through the STEM Graduate Fellowship.

\appendix
\section{Sommerfeld Boundary Conditions}
\label{sec:appendix_sommerfeld}

Sommerfeld boundary conditions (BCs)---also referred to as radiation
or transparent boundary conditions---aim to enable outgoing wave
fronts to pass through the boundaries of a domain with
minimal reflection. Sommerfeld BCs typically assume that
for large values of $r$ any given field $f=f(t,r)$ behaves as an
\emph{outgoing} spherical wave, with an asymptotic value $f_0$ as
$r\to\infty$. Following \newrad~\cite{Alcubierre:2002kk},
our {\it ansatz} for $f(t,r)$ on the boundary takes the form
\begin{equation}
\label{eq:ansatz_sommerfeld}
  f = f_0 + \frac{w(r-ct)}{r} + \frac{C}{r^n}\,,
\end{equation}
where $w(r-ct)/r$ represents an outgoing wave that solves the
wave equation in spherical symmetry,\footnote{That is, $\partial_t^2 (rw)-c^2 \partial_r^2 (rw) = 0$.} and $C$ is a
constant. The $1/r^n$ correction term encapsulates higher-order
corrections with $n>1$ fall-off.

We follow the hyperbolic relaxation
method of~\cite{Ruter:2017iph} and \newrad, and apply Sommerfeld
boundary conditions not to $f$ directly, but to
$\partial_t f$:
\begin{equation}
  \partial_{t}f = -\wavespeed \frac{w'(r-ct)}{r}\,.\label{eq:sommerfeld_diff_basic}
\end{equation}
To better understand the $w'(r-ct)$ term, we compute the
radial partial derivative of $f$ as well:
\begin{equation}
  \partial_{r}f = \frac{w'(r-ct)}{r} - \frac{w(r-ct)}{r^2} - n \frac{C}{r^{n+1}}\,.\label{eq:sommerfeldwaveprime}
\end{equation}
Solving~\eqref{eq:sommerfeldwaveprime} for $w'(r-ct)$ and
substituting into~\eqref{eq:sommerfeld_diff_basic} yields
\begin{equation}
  \partial_{t}f = -\wavespeed \left[\partial_r f + \frac{w(r-ct)}{r^2} + n \frac{C}{r^{n+1}} \right]\,.\label{eq:sommerfeld_unknown_woverr2}
\end{equation}
To take care of the (as-yet) unknown $w(r-ct)/r^{2}$ term, notice that
our \emph{ansatz} \eqref{eq:ansatz_sommerfeld} implies
\begin{equation}
  \frac{w(r-ct)}{r^2} = \frac{f - f_0}{r} - \frac{C}{r^{n+1}}\,,
\end{equation}
which when inserted into~\eqref{eq:sommerfeld_unknown_woverr2} yields
\begin{equation}
  \partial_{t}f = -\wavespeed \biggl[\partial_r f + \frac{f - f_0}{r}\biggr] + \frac{k}{r^{n+1}}\,,
\end{equation}
where $k=-\wavespeed C(n-1)$ is just another constant. Thus we have
derived the desired boundary condition
\begin{equation}
  \partial_{t}f = -\frac{\wavespeed}{r} \bigl[r \partial_r f + (f - f_0)\bigr] + \frac{k}{r^{n+1}}\,.
  \label{eq:sommerfeld_core_eq}
\end{equation}

To generalize~\eqref{eq:sommerfeld_core_eq} to arbitrary curvilinear
coordinate systems $x^{i}_{\rm Curv}$, we make use of the chain rule
\begin{equation}
  \frac{\partial f\bigl(x^{i}_{\rm Curv}\bigr)}{\partial r} = \biggl(\frac{\partial x^{i}_{\rm Curv}}{\partial r}\biggr)\biggl(\frac{\partial f}{\partial x^{i}_{\rm Curv}}\biggr)\,,
\end{equation}
which can be plugged into~\eqref{eq:sommerfeld_core_eq} to give
us~\eqref{eq:hyprelax_sommerfeld}
\begin{equation}
  \partial_{t}f = -\frac{\wavespeed}{r} \left[r \frac{\partial x^i_{\rm Curv}}{\partial r} \frac{\partial f}{\partial x^{i}_{\rm Curv}} + (f - f_0)\right] + \frac{k}{r^{n+1}}\, .
\end{equation}

Returning to the original {\it ansatz}
(\eqrefalt{eq:ansatz_sommerfeld}), we would generally expect the
lowest-order correction to be one order higher than the dominant,
$1/r$ falloff. As the correction term in
\eqref{eq:ansatz_sommerfeld} has $1/r^n$ falloff, we therefore set
$n=2$ to obtain our final expression for imposing outgoing radiation
boundary conditions any given field $f$:
\begin{equation}
\label{eq:sommerfeld_final}
  \boxed{\partial_{t}f = -\frac{\wavespeed}{r} \left[r \frac{\partial x^i_{\rm Curv}}{\partial r} \frac{\partial f}{\partial x^{i}_{\rm Curv}} + (f - f_0)\right] + \frac{k}{r^{3}}}\ .
\end{equation}

Regarding numerical implementation of this expression a couple of
subtleties arise. First, note that $\partial x^{i}_{\rm Curv}/\partial r$ may be impossible to
compute analytically, as the spherical radius $r$ is generally easy to
write in terms of the curvilinear coordinates $x^{i}_{\rm Curv}$, but
not its inverse $x^{i}_{\rm Curv}(r)$.

To address this, for all coordinate systems $x^{i}_{\rm Curv}$
implemented in \nrpy, the function
\begin{equation}
  x^{i}_{\rm Sph} = \biggl(r(x^{i}_{\rm Curv}),\theta(x^{i}_{\rm Curv}),\phi(x^{i}_{\rm Curv})\biggr)\,,
\end{equation}
is explicitly defined. If we define the Jacobian
\begin{equation}
  J^{j}_{\ i} = \frac{\partial x^{j}_{\rm Sph}}{\partial x^{i}_{\rm Curv}}\,,
\end{equation}
and use \nrpy functions to invert this matrix, we obtain exact
expressions for the inverse Jacobian matrix, which encodes
$\partial x^{i}_{\rm Curv}/\partial x^{j}_{\rm Sph}$:
\begin{equation}
  \bigl(J^{-1}\bigr)^{i}_{\ j} = \frac{\partial x^{i}_{\rm Curv}}{\partial x^{j}_{\rm Sph}}\,.
\end{equation}
From this, we can express $\partial x^i_{\rm Curv}/\partial r$
exactly for any curvilinear coordinate system implemented within \nrpy.

The second subtlety lies in formulating a way to approximate $k$. If the function
$f$ represented only an outgoing spherical wave, then it would exactly satisfy
the advection equation
\begin{equation}
  \biggl[\frac{\partial f}{\partial t}\biggr]_{\rm adv} \equiv -\frac{\wavespeed}{r} \bigg[ r\frac{\partial x^{i}_{\rm Curv}}{\partial r}\frac{\partial f}{\partial x^{i}_{\rm Curv}} + (f - f_0) \bigg]\,,
  \label{eq:sommerfeld_advection}
\end{equation}
which is identical to \eqref{eq:sommerfeld_final} but with $k=0$.

Next consider an interior point $r_{\rm int}$ directly adjacent to the outer
boundary. Then, $f(r_{\rm int})$ approximately satisfies \emph{both} the
time evolution equation (e.g. \eqrefalt{eq:two_punctures_rfm_system}),
and the advection equation~\eqref{eq:sommerfeld_advection}. We
compute $\partial_t f(r_{\rm int})$ for a given field $f$ directly
from evaluating the corresponding right-hand side of
\eqref{eq:generic_hyprelaxation}, and
$[\partial_t f]_{\rm adv}(r_{\rm int})$ from
\eqref{eq:sommerfeld_advection}. The
difference of these two equations yields the departure from the
expected purely outgoing wave behavior at that point
$k/r_{\rm int}^{n+1}$. From this we can immediately extract $k$:
\begin{equation}
  \boxed{k = r_{\rm int}^{3} \left(\frac{\partial f}{\partial t} - \left[\frac{\partial
      f}{\partial t} \right]_{\rm adv}\right)_{\rm int}}\ ,
  \label{eq:sommerfeld_expression_for_k}
\end{equation}
where again we impose $n=2$.

Our numerical implementation of Sommerfeld BCs evaluates $\partial
f/\partial x^{i}_{\rm Curv}$ in \eqref{eq:sommerfeld_advection} using either centered or fully upwinded finite-difference derivatives
as needed to ensure finite-difference stencils do not reach out of
bounds. Unlike \newrad, which only implements second-order
finite-difference derivatives for
$\partial f/\partial x^{i}_{\rm Curv}$, our implementation supports
second, fourth, and sixth-order finite differences.

We validated this Sommerfeld boundary
condition algorithm against \newrad for the case of a scalar wave
propagating across a 3D Cartesian grid, choosing second-order
finite-difference derivatives in our algorithm. We achieved
roundoff-level agreement for the wave propagating toward each of the
individual faces.


\section{Spatially-dependent wavespeed and relaxation-wave-crossing time}
\label{sec:appendix_wavespeed}
\begin{figure}[!h]
	\begin{tabular}{c}
		\includegraphics[width=0.5\textwidth,clip]{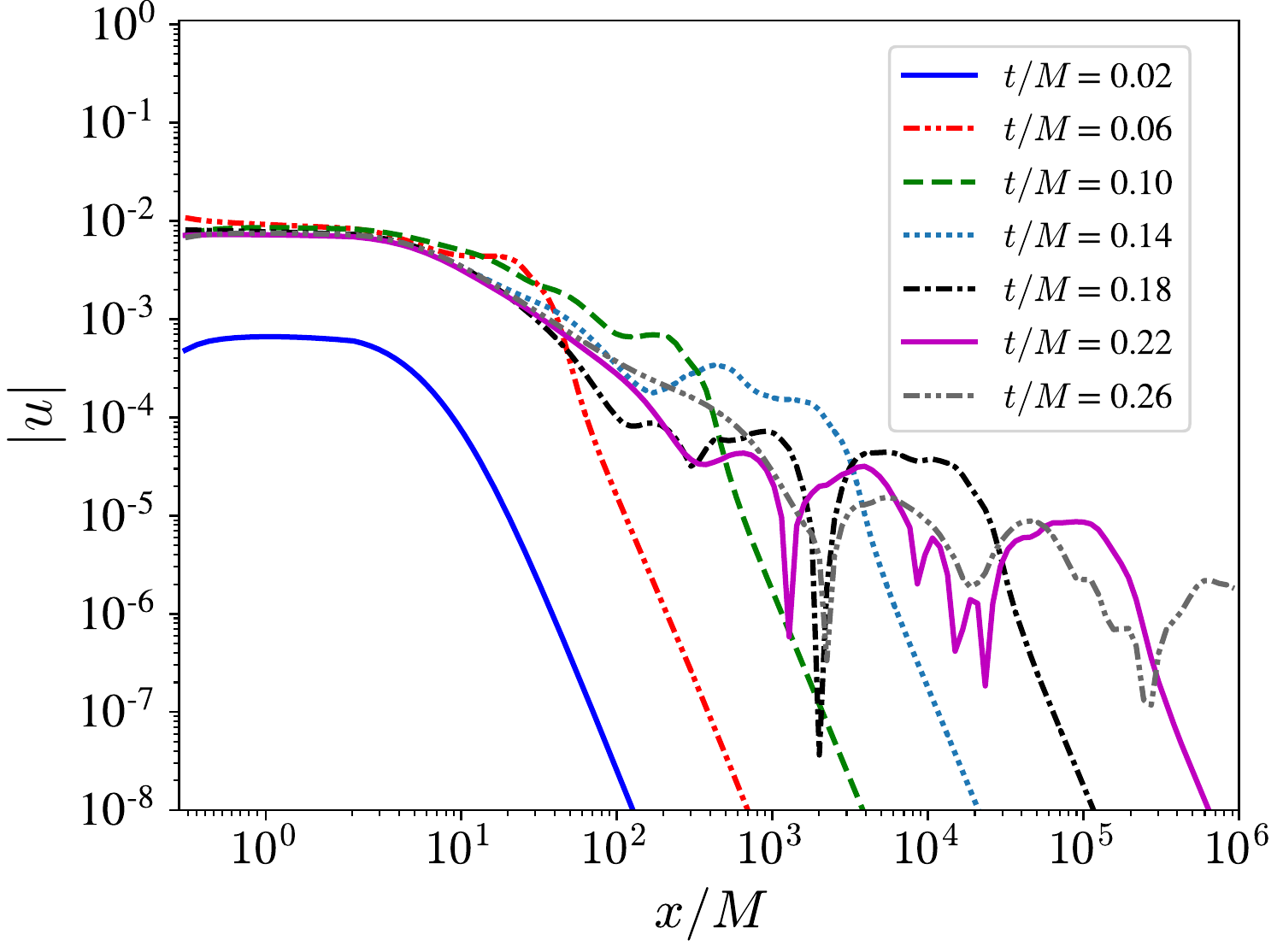}\\
		\includegraphics[width=0.5\textwidth,clip]{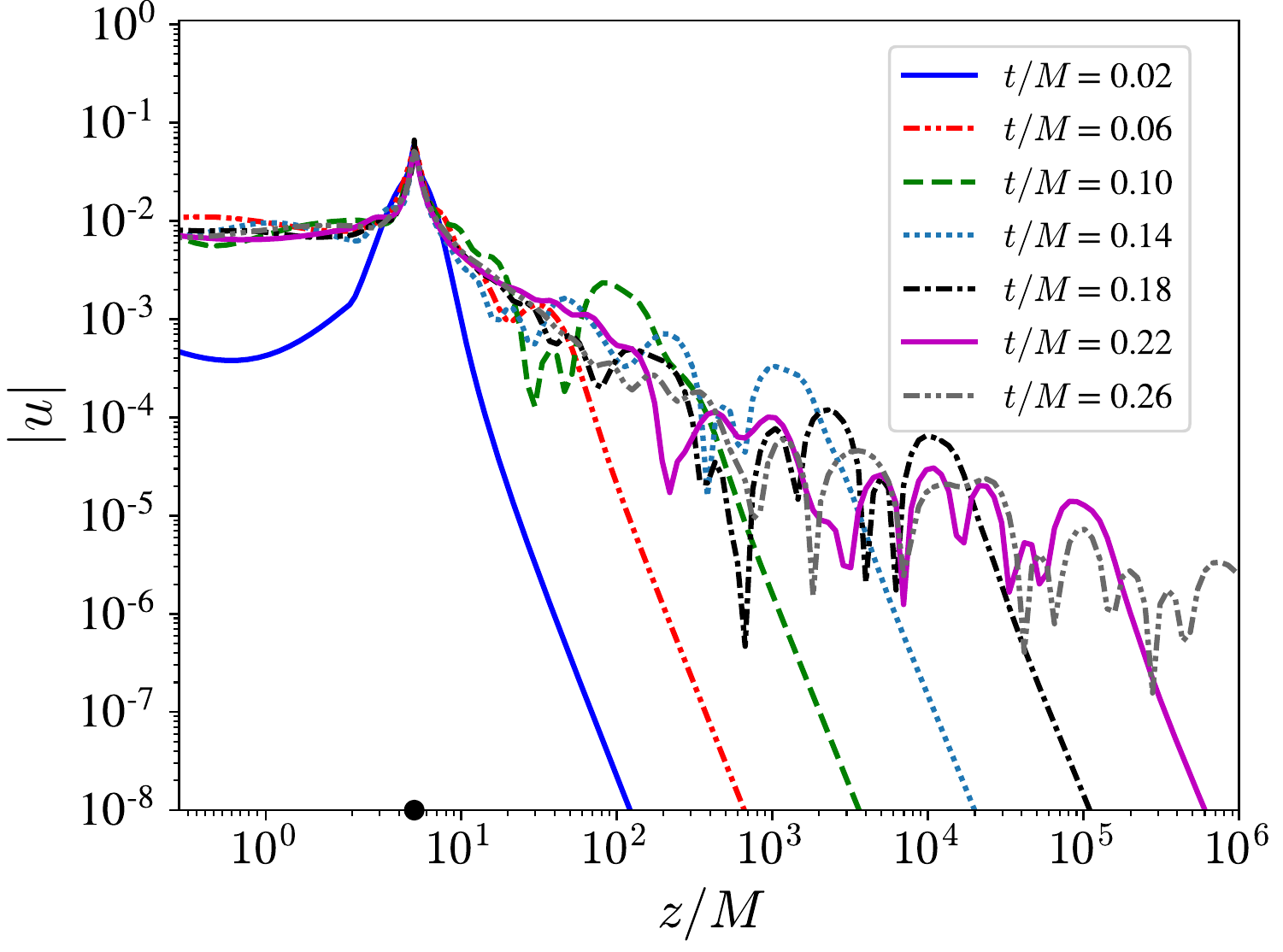}
	\end{tabular}
	\caption{
 	\textbf{Top}: Propagation of $u$ along gridpoints closest to the $x$-axis during the first relaxation-wave-crossing time for the 3D ID.\
	\textbf{Bottom}: Same as top panel, except for gridpoints closest to the $z$-axis.
	\label{fig:wave_propagation_axes}}
\end{figure}

At every point in the domain we compute the smallest proper distance
between neighboring points, $\dS_{ \rm{min} }$, and define a local
wavespeed that is proportional to the grid spacing as
\begin{equation}
    \label{eq:varying_wavespeed}
    c(\vec{x}) = \mathcal{C}_0 \frac{\dS_{\rm{}min} (\vec{x}) }{\dt} \,,
\end{equation}
where $\dt$ is the time step used by the time integrator, and
$\mathcal{C}_0$ is the CFL factor.

For the sake of readability, we repeat here the relationship between \SinhSymTP coordinates and Cartesian coordinates,
\begin{align}
\label{eqn:cords_appendix}
  x &= \tilde{r}\sin(\xxone)\cos(\xxtwo)\,,\nonumber\\
  y &= \tilde{r}\sin(\xxone)\sin(\xxtwo)\,,\\
  z &= \left(\tilde{r}^{2} + \bScale^{2}\right)^{1/2}\cos(\xxone)\,.\nonumber
\end{align}
To establish the time scales associated with the propagation of relaxation waves, we
calculate the relaxation-wave-crossing time along the
$x$-axis---$\left(\xxone,\xxtwo\right)=\left(\pi/2,0\right)$ in
\eqsref{eqn:cords_appendix}---and the $z$-axis. \nrpyell adopts
topologically Cartesian, cell-centered grids with inter-cell spacing of $\dxxzero$,
$\dxxone$, and $\dxxtwo$ in the $\xxzero$, $\xxone$, and
$\xxtwo$-directions, respectively. Because of this, the closest points
to the $x$-axis are obtained by setting $\xxone=(\pi+\dxxone)/2$ and
$\xxtwo=\dxxtwo/2$. Likewise, the closest points to the $z$-axis are
given by setting $\xxone = \dxxone/2$ and $\xxtwo$ to any fixed
value. The relaxation-wave crossing time along the $x$-axis is given by
\begin{align}
    \label{eq:RCT_x-axis_1}
    T_{\rm{RC}}^{(x)} &= \int \frac{dx}{c(x)} \
                    = \int_{\xxzero=0}^{1} \frac{f(\xxzero) d \xxzero}{c^{(x)}(\xxzero)} \,,
\end{align}
where
\begin{equation}
  f(\xxzero) = \partial_{\xxzero}x(\xxzero,\xxone,\xxtwo)\Big|_{\xxone=(\pi+\dxxone)/2,\xxtwo=\dxxtwo/2}\,,
\end{equation}
and $c^{(x)}$ is the wavespeed along the $x$-axis. For the choice of grid
parameters described in Sec.~\ref{sec:numerical_implementation}, the
smallest grid spacing along the $x$-axis is given by the proper distance
in the direction of the $\xxone$ coordinate, so that
\begin{equation}
	\label{eq:wavespeed_x-axis}
    c^{(x)}(\xxzero) = \mathcal{C}_0 \frac{\dSone(\xxzero)}{\Delta t} \
                  = \frac{\mathcal{C}_0 \dxxone}{\Delta t}\hone(\xxzero) \,,
\end{equation}
where $\hone(\xxzero) = \sqrt{\tilde{r}^2(\xxzero) + \bScale^{2} \cos^2(\dxxone/2)}$
is the scale factor. Substituting \eqref{eq:wavespeed_x-axis} in
\eqref{eq:RCT_x-axis_1} and integrating yields

\begin{eqnarray}
T_{\rm{RC}}^{(x)} &=&  \frac{\Delta t}{\mathcal{C}_0 \dxxone} \
            \arctanh \left( \frac{\sinhA}{ \sqrt{\sinhA^2 + \bScale^{2} \cos^2(\dxxone/2)} } \right) \nonumber \\
            &&\times\sin(\dxxone/2) \cos(\dxxtwo/2) \,.
\label{eq:RCT_x-axis_2}
\end{eqnarray}

Similarly, the wavespeed along the $z$-axis is given by
\begin{equation}
\label{eq:wavespeed_z-axis}
c^{(z)}(\xxzero) = \mathcal{C}_0 \frac{\dStwo(\xxzero)}{\Delta t} \
= \frac{\mathcal{C}_0 \dxxtwo}{\Delta t}\htwo(\xxzero) \,,
\end{equation}
where $\htwo(\xxzero) = \tilde{r}(\xxzero) \sin(\dxxone/2)$ is the scale factor in the direction of $\xxtwo$. Thus, the relaxation-wave-crossing time along the $z$-axis can be computed as
\begin{align}
    \label{eq:RCT_z-axis}
    T_{\rm{RC}}^{(z)} &= \int \frac{dz}{c(z)} \
     = \int_{\xxzero=0}^{1} \frac{g(\xxzero) d \xxzero}{c^{(z)}(\xxzero)}  \nonumber \\
    &= \frac{\Delta t}{\mathcal{C}_0 \dxxtwo} \
       \arctanh \left( \frac{\sinhA}{ \sqrt{\sinhA^2 + \bScale^{2}} }
       \right) \cot(\dxxone/2) \,,
\end{align}
where we used
\begin{equation}
  g(\xxzero) = \partial_{\xxzero}z(\xxzero,\xxone,\xxtwo)\Big|_{\xxone=\dxxone/2}\,.
\end{equation}

Plugging in the values of the grid parameters $\sinhA$, $\sinhW$, and
step sizes, we find $T_{\rm{RC}}^{(x)} = 0.248M$ ($0.574M$) and
$T_{\rm{RC}}^{(z)} = 1.28M$ ($1.26M$) for full-3D (axisymmetric) ID.

From \eqref{eq:wavespeed_z-axis} we find that for fixed values of
$\xxzero$ and small angles $\xxone$ the wavespeed along the $z$-axis behaves as
\begin{equation}
  c^{(z)} \sim \tilde{r}(\xxzero)\,\xxone\,\dxxtwo\,.
\end{equation}
Our cell-centered grid has \mbox{$\xxone_{\ione}=(\ione + 1/2)\dxxone$} with
\mbox{$\ione=0,1,2,\ldots,\None$}, and thus the wavespeed quickly
increases as we move away from the axis. Such rapidly-moving signals slightly off the $z$-axis are not considered in our analytic estimates, yet influence
the RCT so that in our full-3D simulations we observe the same value of
$T_{\rm RC} \approx 0.25M \approx T_{\rm RC}^{(x)}$ along {\it both} axes, as shown in
Fig.~\ref{fig:wave_propagation_axes}. Thus we use $T_{\rm RC}=T_{\rm RC}^{(x)}$ in all figures.

\bibliography{references}

\end{document}